\documentclass[a4paper,twoside,10pt]{article}
\usepackage{amsmath}
\usepackage{amssymb}
\usepackage{amsfonts}
\usepackage{graphicx}
\usepackage{epstopdf}

\newcommand{\Ron}{R_{\mathrm{on}}}
\newcommand{\Roff}{R_{\mathrm{off}}}
\newcommand{\Rcon}{R_{\mathrm{con}}}

\title{Which Memristor Theory is Best for Relating Devices Properties to Memristive Function?}
\author{Ella M. Gale, Benjamin de Lacy Costello and Andrew Adamatzky}

\begin{document}
\maketitle


\begin{abstract}
There are three theoretical models which purport to relate experimentally-measurable or fabrication-controllable device properties to the memristor's operation: 1. Strukov et al's phenomenological model; 2. Georgiou et al's Bernoulli rewrite of that phenomenological model; 3. Gale's memory-conservation model. They differ in their prediction of the effect on memristance of changing the electrode size and factors that affect the hysteresis. Using a batch of TiO$_2$ sol-gel memristors fabricated with different top electrode widths we test and compare these three theories. It was found that, contrary to model 2's prediction, the `dimensionless lumped parameter', $\beta$, did not correlate to any measure of the hysteresis. Contrary to model 1, memristance was found to be dependent on the three spatial dimensions of the TiO$_2$ layer, as was predicted by model 3. Model 3 was found to fit the change in resistance value with electrode size. Simulations using model 3 and experimentally derived values for contact resistance gave hysteresis values that were linearly related to (and only one order of magnitude out) from the experimentally-measured values. Memristor hysteresis was found to be related to the ON state resistance and thus the electrode size (as those two are related). These results offer a verification of the memory-conservation theory of memristance and its association of the vacancy magnetic flux with the missing magnetic flux in memristor theory. This is the first paper to experimentally test various theories pertaining to the operation of memristor devices.
\end{abstract}

\maketitle

\section{Introduction}

\begin{figure}[htbp!]
 \centering
 \includegraphics[width=5in]{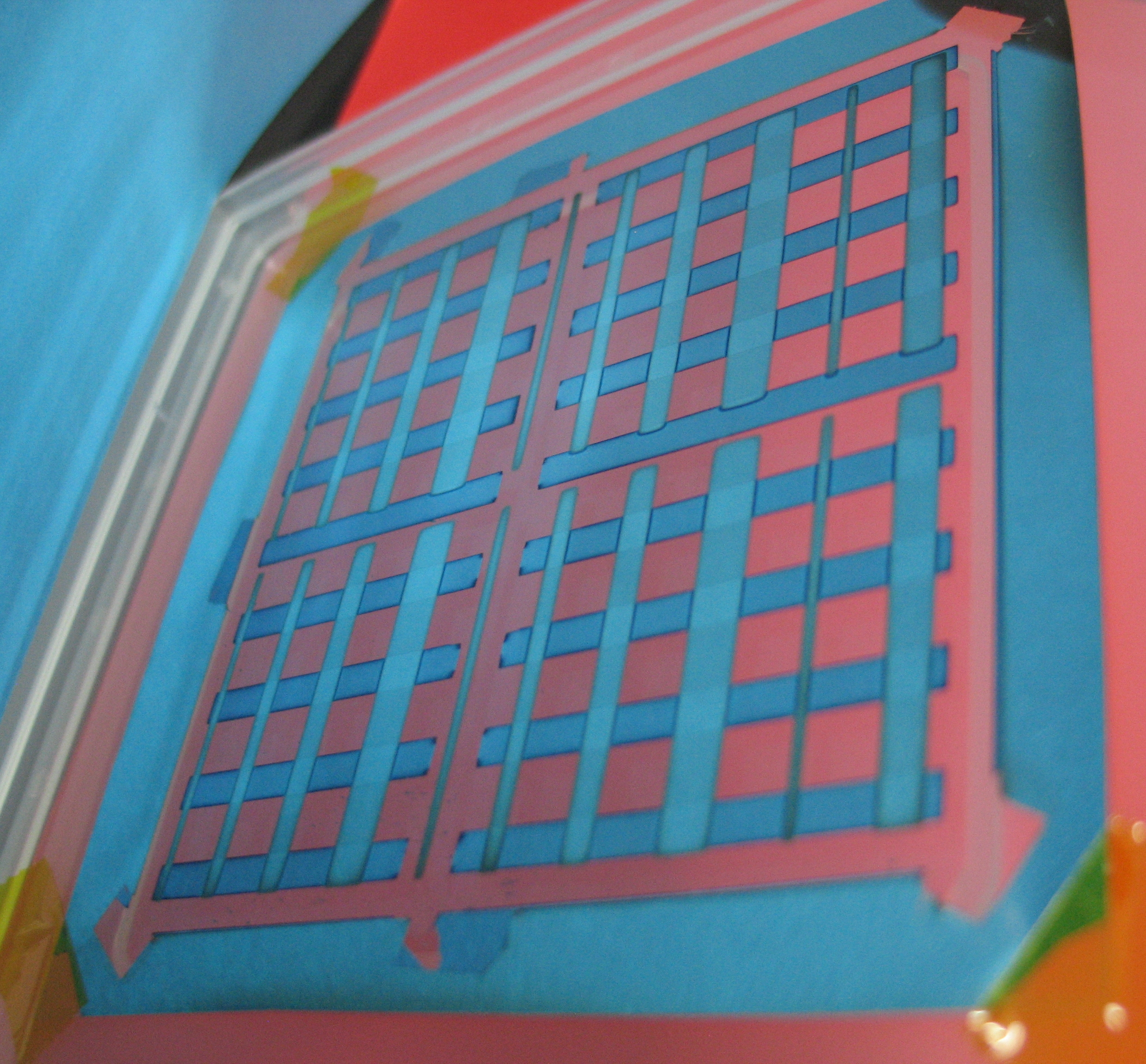}
 \caption{A batch of 64 flexible memristors of different sizes used in the experiments. This photo has been shot with red coloured paper backing and blue paper reflector for clarity, the TiO$_2$ gel is transparent and the aluminium electrodes are silver-coloured and highly reflective.}
 \label{fig:DSeriesMemristors}
\end{figure}

The development of modern science has had a huge impact on humanity's progress only because experimental verification gives a method for sorting ideas by their accuracy in describing real world phenomena. When developing a novel technology design rules and methodologies are required, whether they are sourced from an abstract theory of what should be or via empirical `rules of thumb' arsing from experimental observations. There are three main memristor theories~\cite{15,224,F1} which model memristor devices and the inventors of two of these~\cite{F1,224} have suggested that their theories will aid those designing and building memristor devices. In this paper we shall experimentally test these theories and these claims. 

The memristor joins the resistor, capacitor and inductor as the 4$^{\mathrm{th}}$ fundamental circuit element~\cite{14}. It has been suggested that the memristor could allow us to build smaller, more energy-efficient computers, less-volatile, more-resilient computer memory, novel, transistor-less computers, and possibly neuromorphic computers which can process information in a brain-like way~\cite{15}. To build these things we must understand the memristor. 

The memristor is defined via the constitutive relation between charge, $q$ and magnetic flux, $\varphi$, as given by

\begin{equation}
d \varphi = M(q(t)) dq \: ,
\label{eq:M}
\end{equation}

where the memristance, $M$, is only dependent on time due to its dependence on $q$~\cite{14}. The memristor is the first (possibly the only if the memcapacitor and meminductor are considered to be hybrid) fundamental non-linear circuit element. Nonlinear electronics are known to give rise to rich dynamics and possibly a wider range of uses than linear electronics~\cite{Chua1969}. Although the memristor is defined by its constitutive relation between $q$ and $\varphi$, it is theoretically predicted to be have a frequency-dependent pinched hysteresis loop~\cite{14,279} in $V-I$ space (note that open-loop versions have been theoretically postulated~\cite{93,276}).  

This definition of the memristor was published in 1971~\cite{14}, but it was not related to a real world device until 2008, when Strukov et al announced that they had made the memristor~\cite{15}. They may not have been the first because many ReRAM (Resistive Random Access Memory) devices were made in the interim, and the possibility that all these devices~\cite{119} or a subset of them~\cite{M1} may be memristors has been postulated. 

Most memristors are made from metal oxides sandwiched between metal electrodes with TiO$_2$ being a popular choice, either deposited via atomic deposition (e.g.~\cite{15}), or solution-processed~\cite{28}, however they have been made from other materials such as chalcengenocide~\cite{104}, Silver nanowires~\cite{71} and conducting polymers~\cite{5,88}. TiO$_2$ memristors are thought to work by the material inter-converting between the stoichiometric (TiO$_2$) and doped (TiO$_{(2-x)}$) forms (although an electrochemical mechanism has been formulated~\cite{94}). The precise structure of the doped form is disputed, with suggestions of Magn\'eli phases~\cite{165}, phase-transition~\cite{155,136}, Ti$_4$O$_7$ conducting channels~\cite{154,168}. The memristor's operational mechanisms in different devices is the subject of much ongoing experimental work~\cite{RevMemReRAM}. Nonetheless, the simplified theoretical model of the memristor as a space-conserving variable resistor as put forward in~\cite{15} has provided an abstraction for all these devices. In this model a boundary, $w(t)$, between the stoichiometric and doped forms of TiO$_2$ moves under the influence of voltage. This causes a change in resistance of the device due to the changing volumes of doped and stoiciometric forms. This description of the memristor device has been used in the three theoretical models discussed in this paper.

The first model of memristor devices which included real world measurables was Strukov's phenomenological model~\cite{15} and came from writing equations of motion for $w(t)$. This model is considered by them a `toy' model not fit for device modelling~\cite{289}, however, due to its apparent simplicity, it has been used in many memristor modelling and simulation papers (some typical examples but by no means an exhaustive list are~\cite{63,65,66,73,78,83}). Furthermore model 1 did not intrinsically take account of the boundaries, which has resulted in the creation of window functions to correct this. In this work we do not consider window functions to be a new rewrite of the theory, but rather modifications to improve the accuracy when modelling the device close to the limits of $w$. As window functions can be applied on top of any theory based on the concept of a space-conserving time-varying resistor, we shall ignore them in our tests here, but suggest that they could be applied to improve accuracy in critical situations. 

The second model of the memristor is a rewrite of the first by Georgiou et al~\cite{224} to yield new information. This required using Bernoulli equations to convert the non-linear equations of motion to a linearised form with known solutions which allow the prediction of hysteresis magnitude (this is not the only paper to investigate the hysteresis of a memristor, for a different approach see~\ref{281,282}, but it is the only that contains an analytical expression of hysteresis). 

The third model of the memristor~\cite{F1} is not based on Strukov's `toy' model, but instead seeks the memristance by calculating the $\varphi(t)$ arising from $q(t)$ by using magnetostatic descriptions of the vacancy current and equation~\ref{eq:M}. This model is not phenomenological but is instead grounded in electromagnetic and circuit theory.

In order to differentiate between these theories via experimental measurements, we first need to examine the theories in depth to tease out the testable differences between them (section~\ref{sec:Theory}). A key difference between the theories relates to their predictions on the effect of changing the electrode width, so we fabricated a batch of 64 memristors with different sized electrodes to test this as shown in figure~\ref{fig:DSeriesMemristors}.

\section{Theoretical Background~\label{sec:Theory}}

All three models use a different set of experimental measurables and fabrication parameters, thus we shall outline what these are before moving on to a summary of the three models and how they might be tested. 

\subsection{Relevant parameters}

The aim of the following three theories is to relate fabrication and material parameters to the device's memristive response, and so in this section we shall briefly introduce and discuss the possible relevant parameters. The first set of parameters are the material properties, so called because to change their value we must change which memristive material we choose to use (e.g. by switching to another metal oxide). The material properties are: the ion mobility of the vacancies, $\mu_v$, and the resistivities of the material in its off and on states, $\rho_{\mathrm{off}}$ and $\rho_{\mathrm{on}}$ respectively. 

The set of parameters which can be altered by changing how the device is fabricated are the fabrication parameters and these are $D$, $E$, $F$, $\Ron$ and $\Roff$. The dimensions of the memristive material is included via $D$ the semi-conductor thickness (i.e. the thickness of the TiO$_2$ sol-gel layer) and the widths of the bottom and top electrodes which are $E$ and $F$ respectively (see figure 1 in~\cite{F1}). $D$, $E$ and $F$ are the limits of the semi-conductor volume in the cardinal directions $x$, $y$ and $z$. $\Ron$ and $\Roff$ are the resistance values of the memristor in its fully switched on and fully switched off states respectively: these values are derived from $\rho_{\mathrm{off}}$, $\rho_{\mathrm{on}}$, $D$, $E$ and $F$ because we are modelling the memristor as a space-conserving variable resistor made of a material that can interconvert between two different resistivity forms. 

Finally we have the parameters which are altered by the specific experiment we chose to run: $L$, $V_{\mathrm{max}}$, $\omega_0$ and $R_0$. The electric feild across the memristor is $L$ (as calculated by $V/D$), $V_{\mathrm{max}}$ is the maximum voltage (i.e. the amplitude of the voltage waveform for a.c. voltage supply), $\omega_0$ is the frequency of the voltage waveform and $R_0$ is the starting resistance, namely the resistance the memristor has at the start of the experiment. 

\subsection{Model 1: Phenomenological Model}

If the memristor is modelled as a space conserving variable resistor then that means that as one part, say the TiO$_2$ increases, the other part, in this case the TiO$_{(2-x)}$, decreases. Strukov et al~\cite{15} chose to model this as a 1-dimensional system, where the movement of $w(t)$ varies between 0 and $D$ and is simply a mixing coefficient of what proportion of the resistance is due to which phase, thus they wrote~\cite{15}:

\begin{equation}
M(t) = \frac{w(t)}{D} \Ron + \left( 1 - \frac{w(t)}{D} \right) \Roff \: ,
\label{eq:R(t)}
\end{equation}

where the first term describes the memristance due to the amount of the memristor which is in doped form and the second describes the memristance due to the amount of material in the stoichiometric form. This equation has one variable ($w$) and is spatially 1-dimensional as it is only concerned with the progress of the boundary along a single direction: we shall take this direction as the $x$ axis in all our models. From equation~\ref{eq:R(t)} they then derived (for a critique of this derivation see~\cite{F1}) the following equation~\cite{15}:

\begin{equation}
M(q) = \Roff \left( 1 - \frac{\mu_v \Ron}{D^2} q(t) \right) \:. 
\end{equation}

(Note that they throw away a small term that is quadratic in $\Ron$ which some researchers include). As a result of this equation, it was erroneously claimed memristors had to be nanoscale because the $D^2$ term needed to be small~\cite{15} for there to be measurable differences between the two states: this supposition has been contradicted by the mesoscale and microscale memristors have now been reported~\cite{236,280,29,M0} and it was also the basis for the supposition~\cite{15} that memristance was only due to the thickness ($D$) of the semi-conductor layer (and thus 1-dimensional) despite the fact that 1-dimensionality was an assumption of the model. 

As the phenomenological model is only dependent on $D$ we can test the model by seeing if there is a difference in memristor behaviour as a result of changing the size of the electrode widths: $E$ and $F$. According to the phenomenological model there should be no effect of this change. 


\subsection{Model 2: Bernoulli Equation Rewrite of the Phenomenological Model}

To get analytical solutions for memristor dynamics, Strukov's model was rewritten as Bernoulli equations in~\cite{224} which allows the calculation of hysteresis from the theory. All the parameters (fabrication and experimental) that were chosen for inclusion in the model were combined into a single `dimensionless lumped parameter', $\beta$, as given by: 

\begin{equation}
\beta = \frac{2 V_{\mathrm{max}}}{\omega_{0} R_{0}^{2}}\mu_{v} \left( \frac{\Ron}{D} \right)^{2} \left( \frac{\Roff}{\Ron} - 1 \right) \: ,  
\label{eq:beta}
\end{equation}

and this parameter is intended to capture the entire dynamical response of a memristor. Georgiou et al investigated the dynamical response of the memristor to three different (a.c.) voltage waveforms: sinusoidal, triangular and bipolar piece-wise linear (BPWL), which is a wedge-shaped waveform in between square and triangular waves (the size of slope relative to the time spent at the maximum voltage was tunable by $m$, see below). A rescaled form of $\beta$ called $\tilde{\beta}$ was defined, and this is the form of the parameter that is actually used to encapsulate the dynamics. This parameter $\tilde{\beta}$ is given by $\tilde{\beta} = s \beta$, where $s$ refers to the set of scale factors $\{ s_s, s_t, s_b \}$ for the sinusoidal ($s$), triangular ($t$) and BPWL ($b$) waveforms and these are given by: $s_s = 2$, $s_t = \frac{\pi}{2}$ and $s_b = \pi - 2 \pi m$ where $m$ relates to the slope of the BPWL. In this paper we will refer to set of values, $\{ \beta \}$, for the three waveforms as $\beta_s$ for sinusoidal, $\beta_t$ for triangular and $\beta_b$ for BPWL, and similarly for $\tilde{\beta}$. 

Georgiou et al also reported analytical expressions for the hysteresis, $H$, and the scaled hysteresis, $\bar{H}$, for the triangular, ($H_t$ and $\bar{H}_t$) and BPWL ($H_b$ and $\bar{H}_b$) waveforms respectively. The scaled hysteresis is calculated relative to the work done by a resistor of $R_0$. The measures for sinusoidal hysteresis, $H_s$ and $\bar{H}_s$ are not analytically solvable but was numerically found to lie in between the results for triangular and BPWL waveforms. Although not explicitly stated in~\cite{224}, $H_t$ and $H_b$ seem to have been chosen as they can provide upper and lower limits for the values of $H_s$ (which is the most useful waveform) and the sinusoidal $I-V$ curves also fell in between the triangular and BPWL $I-V$ curves.


From a thorough reading of~\cite{224}, we get the following theoretical predictions which are experimentally testable: 
\begin{itemize}
\item Test 1: $ H_t < H_s <H_b$, the hysteresis of a memristor under sinusoidal voltage excitation should be larger than one under triangular voltage waveform and smaller than one undergoing BPWL waveform, a specific example is given in~\cite{224} for $\tilde{\beta}=0.9$.
\item Test 2: $\tilde{\beta} \propto \bar{H}$, the hysteresis should be related to $\tilde{\beta}$, specifically it should follow a monotonically increasing curve as given in figure 3a in~\cite{224}.
\end{itemize}

Assuming that the reader has read the abstract, they will know that a calculation of $\tilde{\beta}$ does not help predict the value of $H$, therefore several other tests were undertaken in order to elucidate if the Bernoulli rewrite could offer useful information for experimentalists. These tests are:
\begin{itemize}
 \item Test 3: $H \propto \omega_0$ and $\bar{H} \propto \omega_0$, or does the hysteresis depend on the frequency?
\item Test 4: $\bar{H}$(theory)$\propto \bar{H}$(experiment) and $H$(theory)$\propto H$(experiment).
\end{itemize}
Test 4 was done to see if the hysteresis values were perhaps qualitatively correct, in the hope that an experimentally measured fitting parameter could allow the prediction of hysteresis based on this model.


\subsection{Model 3: Memory-Conservation Theory}

The memory-conservation theory~\cite{F1} models the memristors as space-conserving variable memristors from the point of view of the memory property (the vacancies) and arrives at a measurable and testable model that fits the constitutive definition for the memristor (the other two models lack this quality due to the lack of a relation for magnetic flux). This theory gives a fundamentally different equation for the memristance which is:
\begin{equation}
M(q) = U X \mu_v P_k (q(t)) \: ,
\label{eq:M_EG}
\end{equation}
where $U$ represents the universal constants given by $U=\mu_0 \diagup 4 \pi$, $X$ is the experimental constants $A L$, where $A$ is the area of a side of the device, and $P_k$ arises from the magnetic field perpendicular to the surface chosen for $A$ (eg. $P_k$ would be $P_z$ for $A=DE$). $P_k$ is a function of $w(t)$, $D$, $E$ and $F$. The $q$ in this theory refers to the charge arising from the memory property (the vacancies), not the electrons (as in the other two theories) and thus to write the memristance from the point of view of the measured electrons (this is called the memory function, $M_e$) we use
\begin{equation}
M_e = C_M M \: ,
\label{eq:Mem}
\end{equation}
where $C_M$ is an experimentally-determined parameter which relates the memristance as experienced by the vacancies with that experienced by the electrons, and it is related to the memristive material chosen for a device. This is a consequence of the idea that resistivity of a material could be different for different charge carriers~\cite{F1} and thus there should be a conversion between ionic and electronic resistances. 

The memory function describes only the doped part of the memristor. To describe the stoichiometric part we used the conservation function, $\Rcon$, as given by

\begin{equation}
\Rcon = \frac{( D-w(t) ) \rho_{\mathrm{off}}}{E F} \: .
\label{eq:Con}
\end{equation}

The total memristance, $R(t)$, is:
\begin{equation}
R(t) = M_e + \Rcon. 
\end{equation}
When the memristor state is at minimum resistance, $R(t)$ is mostly dependent on the memory function part of the model, $R(t) \sim M_e$; similarly, when $R(t)$ is at it's maximum value, $\Roff$, we can approximate it as $R \approx \Rcon$. We don't know the values of $C_M$ and $\rho_{\mathrm{off}}$ for our devices but we can look at the variation of $\Ron$ with $F$ and attempt to fit it by $M_{e}(F)$ and similarly, we can compare $\Roff(F)$ with $\Rcon(F)$. This would demonstrate that the electrode width is an integral part of memristance as well as test the usefulness of the memory-conservation model.  

This theory is spatially three-dimensional and so we would expect that changing the electrode widths would have an effect on the measured memristance. Thus, the memory-conservation model offers the following experimentally-testable questions:
\begin{itemize}
 \item Test A: $M \propto F$, is the memristance related to $F$?
 \item Test B: $\Ron \propto M_e$, can the ON state resistance be related to the memory function?
 \item Test C: $\Roff \propto \Rcon$, can the OFF state resistance be related to the conservation function?
\end{itemize}

The memory-conservation model doesn't currently offer any analytical expressions for the value of the hysteresis, however it can be used to simulate our devices and get a hysteresis value that way. Using numbers calculated from our data for the 16 different devices, simulations of them were run and the hysteresis calculated as for the experimental data.

\subsection{Titanium Dioxide Sol-gel Memristors}

We used TiO$_2$ sol-gel memristors with a thickness of circa 40nm~\cite{M1}. The memristors were classified into two types: those with an ohmic high resistance state several orders of magnitude above the low resistance state which is similar to the Unipolar Switching (UPS) seen in ReRAM devices (there is some subtleties we are ignoring here, see~\cite{M1} for a more thorough discussion); those with nonlinear high and low resistance states which were around the same order of magnitude which is similar to bipolar switching seen in BPS. Although it has been claimed that UPS ReRAM devices are memristors~\cite{119} and that memristance is a useful theory with little practical relevance in understanding  ReRAM mechanisms~\cite{WasersBook}, we take the position that memristors BPS ReRAM and memristors are probably the same thing. Thus, in this paper, we will only use the BPS-like devices as these are the device that the majority of ReRAM and memristor researchers consider to be the closest to Chua's theory. Note that the Memory-Conservation theory can be extended to include the growth of conducting filaments~\cite{254} which allows a better model of UPS-like memristors.

BPS-like memristors were found to have memristor-like I-V curves, in that they demonstrated pinched hysteresis loops over large voltage ranges~\cite{M1}. However, when run over small voltage ranges, it was found that the curves were pinched but did not cross zero, which does not fit the original definition~\cite{14} but which does fit later work~\cite{93,276}. Work is currently underway to pin down what causes this effect.

\section{Methodology}

\subsection{Testing the effect of size}

As the phenomenological model predicts no effect of changing the values of $E$ and $F$ and the memory-conservation model does, we can simply make and measure memristors of different sizes. To do this we made a batch of sol-gel memristors with one electrode, $E$, set to our standard width, 4mm, and the other, $F$, set to 1, 2, 3, 4 or 5mm. 64 memristors in total were made and 16 were classified as BPS-like and suitable for the test.

Even if the Strukov model does not predict any effect of the values of $E$ and $F$, it could be argued that there will be an effect of increasing electrode size due to the fact the active area has been increased. For this reason, the memory-conservation model was fitted to the maximum and minimum values, as described below, in order to demonstrate its effectiveness. If the electrode size has an effect on the time-varying resistance, then it should be possible to fit and quantify that effect.

The I-V curves of the virgin devices were run over a $\pm$0.5V and they were plotted to see if there was an effect of electrode size.

\subsection{Measuring the Hysteresis}

We follow Georgiou et al's definition for the theoretical hysteresis~\cite{224}: the difference between the upper (2 and 4) and lower (1 and 3) branches in the integration of the instantaneous power consumed by the device over the course of an input cycle or
\begin{equation}
W_A = \int_{t_{0}}^{t_{0}+T/4} i(t) v(t) dt \:,
\end{equation}
where the final value of the hysteresis is given as $H = (W_2 + W_3) - (W_1 + W_4)$, which is simply the difference in work between the upper and lower parts of the I-V curve~\cite{224} (the I-V curve has been split into 4 branches as 1: 0V$\rightarrow +V_{max}$; 2: $+V_{max} \rightarrow$ 0V; 3: 0V$\rightarrow -V_{max}$ and 4:$-V_{max} \rightarrow$ 0V and $A_0$ is the starting point for branch A) and $T$ is the period.

To calculate the experimental equivalent, we first calculate the power per branch in a descretized fashion (as we only have information at every data point). For branch A the work is
\begin{equation}
W_A = \sum_{n=1}^{n=1+N/4} I(n) V(n) \Delta t \: ,
\label{eq:Work}
\end{equation}
where $I(n)$ and $V(n)$ are the current and voltage values at that data point, and $\Delta t$ the time between measurements and $N$ is the total number of measured points. The measured hysteresis is then calculated as for the theory by taking the difference between the work for the upper and lower branches of the I-V curve.  

\subsection{Frequency and waveform effects to Test the Bernoulli Equation Rewrite of the Phenomenological Model}

Calculating the theoretical hysteresis values involves several steps. To get hysteresis measure for a specific $\tilde{\beta}$ we need to run each waveform at a different frequency. For each value of $\tilde{\beta}$ we first calculate the unscaled $\beta$ values for the three waveforms ($\beta_s, \beta_t, \beta_b$). To get the required frequencies, { $\omega_0(s), \omega_0(t), \omega_0(b)$ }, we rearrange equation~\ref{eq:beta} and substitute $\beta$ with $\{ \beta_s, \beta_t, \beta_{b} \}$. These $\beta$ are then substituted in the equations for the scaled hysteresis ($\bar{H}_t$ and $\bar{H}_{b}$) and if we know $R_{0}$ we can calculate the measurable hysteresis, $H$, by $H = \bar{H} R_0$. We can also start with a frequency we want to measure at, and calculate the required $\beta$ and $H$. We used $m=1/20$ for the BPWL waveform as in~\cite{224}.

The actual experimental proceedure is as follows. To get max($\Roff$) and min($\Ron$) for use in the equations we did long-time $N=4008$, $\Delta t=2s$ d.c. $I-t$ measurements by sourcing a constant voltage and measuring the current response (the time chosen was the result of testing different values to ensure the device had fully switched). The current would stabilise to a value~\cite{SpcJ}. From the last point of this graph we calculated the starting $R_0$ from $V(T) \diagup I(T)$.  The values for $\Roff , \Ron$ and $R_0$ were substituted into the equations for $\beta$, with $\mu_{v}=1\times10^{-10}cm^{-2}Vs^{-1}$ as in~\cite{15}, $D$=40nm, $V_{\mathrm{max}}$=1V and from this the value of one of the $\{ \tilde{\beta} \}$ and its relevant $\omega_0$ was calculated. These equations were in put into a Mathematica script, which could be reevaluated to get the different output values. The required waveform was run at $\omega_0$ using a Keithley electrometer 2400 set in voltage-sourcing, current-sensing mode where the voltages are read in from MatLab allowing us to program any waveform (and achieve the required low frequencies). A high compliance current is used with no set current range to allow the electrometer to get very accurate measures of the current; this can distort the frequency as the electrometer will take longer per time-step if it has to change current range. Thus, once we know the current range for that device at that run, we recalculate $R_0$, if changed, and reevaluate the equations to get an updated frequency, generally the changes were small but non-negligible. A second run is then done with the current range set to the value found before to reduce the variance on the time-step and give the exact requested frequency. The compliance current to 1mA (i.e. far above what we will be measuring). Note, the Keithley measurement steps are padded by 0.6s, which is the time taken to settle and do the measurement, thus this is taken into account when calculating the correct $\Delta t$ for our desired frequency: the measurement `time' gives an upper limit on the frequency we can use to measure our devices. The output $I-V$ loop is plotted, the work and hysteresis calculated as above (using MatLab scripts) and compared to the theoretical calculations. Therefore, to get a single $\beta$ value's worth of data for the hysteresis graphs requires 6 separate experiments (not including the repeated runs to set the current range).

To test Georgiou's model two studies were undertaken. First we used the data in section~\ref{sec:AveCurves}, which was the virgin runs of 16 devices. Although they were run at the same frequency, the variance in $R_0$ offered a small variance in $\tilde{\beta}$ which could then be compared to the measured hysteresis ($\tilde{\beta}$ covered the range $\approx 0-0.03$). One of the best devices was then picked for the second study which involved repeated runs with the three frequencies (corresponding to the three waveforms) for each value of $\tilde{\beta}$ in the range $0.4 - 0.9$ following the experimental proceedure discussed in the previous paragraph. 

Due to the measurement time limit, all three waveforms are measured at 0.05$\tilde{\beta}$ increments between 0.6 and 1, plus a measurement of the sinusoidal and triangular waveforms at $\tilde{\beta}=0.5$ and the triangular 
waveform at $\tilde{\beta}=0.4$, because each waveform hits this limit at a different point. There are two measurements for $\tilde{\beta}=0.9$ for the sinusoidal and BPWL waveforms, resulting in 12 $\tilde{\beta}$ values tested.

\subsection{Fitting the memory-conservation Model}

The experimentally measured values of $\Roff$ and $\Ron$ were taken from the experiment by taking the resistance at the end of $w_2$ to be $\Roff$ and the resistance corresponding to the highest current state (generally the end of $w_1$) as $\Ron$. Note that these values are both taken from the positive lobe of the I-V curves. We could have taken values from the negative voltage part of the curve, but as our devices are not completely symmetrical, these values give an smaller $\Roff$ and larger $\Ron$ and thus do not correspond to the limits of the device. The starting resistance, $R_0$ was taken as the first resistance measured in the I-V curve, this did change from run to run in the repeats, but not by much. 

When the memristor state is at minimum resistance, $R(t)$ is mostly dependent on the memory function part of the model. As $w \rightarrow D$, $M_e \rightarrow M_e(\mathrm{max})$, $\Rcon \rightarrow \Rcon(\mathrm{max})$ and so we can approximate $\Ron \approx M_e$. Similarly, as $w \rightarrow 0$, $M_e \rightarrow M_e(\mathrm{min})$, and $\Rcon \rightarrow \Rcon(\mathrm{max})$, $R(t)$ is at it's maximum value, $\Roff$, we can approximate it as $\Roff \approx \Rcon$. These approximations are only true if we run the device at the frequency that will allow it to fully switch off and on, $\omega_0$, which will also give the largest hysteresis.

To fit the conservation function, we used the resistivity of the `insulating' material, $\rho_{\mathrm{off}}$, as a fitting parameter in equation~\ref{eq:Con}. For the memory function, the curve of $F$ versus $\Ron$ was fit to $R_{\mathrm{on}} = M_e(C*F)$ rather than $R_{\mathrm{on}} = C_m*M_e(F)$ as the optimisation algorithm had problems with the latter form. $C_m$ is a measure of how the ionic memristance effects the resistance experienced by the electrons, $C$ is that measure recast as a different effective size of the electrode. In other words, the electrode size $F$ `feels' bigger to the electrons as it includes the conversion between ionic and electronic resistance.

Fits were performed using MatLab's unconstrained non-linear optimisation method `fminseach' which is based on the Nelder-Mead Simplex Method~\cite{MatLabfminsearch}. Straight line fits were performed using MatLab's simple fitting function, which gives the norm of the residuals as a measure of goodness of fit.

\subsection{Simulating Using the Memory-Conservation Model}

Simulations were performed in MatLab using scripts written that can calculate the anticipated current from a voltage waveform input using the memory-conservation equations. All values were taken from the experiments, including the fitted values of $\rho_{\mathrm{off}}$ and $C_m$. The hysteresis from the simulations were calculated as if the data was experimental. Simulations were run for the same number of steps as the experiment gathered.  

\section{Results}

\subsection{Does the Electrode Width have an Effect on the Memristance?~\label{sec:AveCurves}}

\begin{figure}
	\centering
		\includegraphics[width=5in]{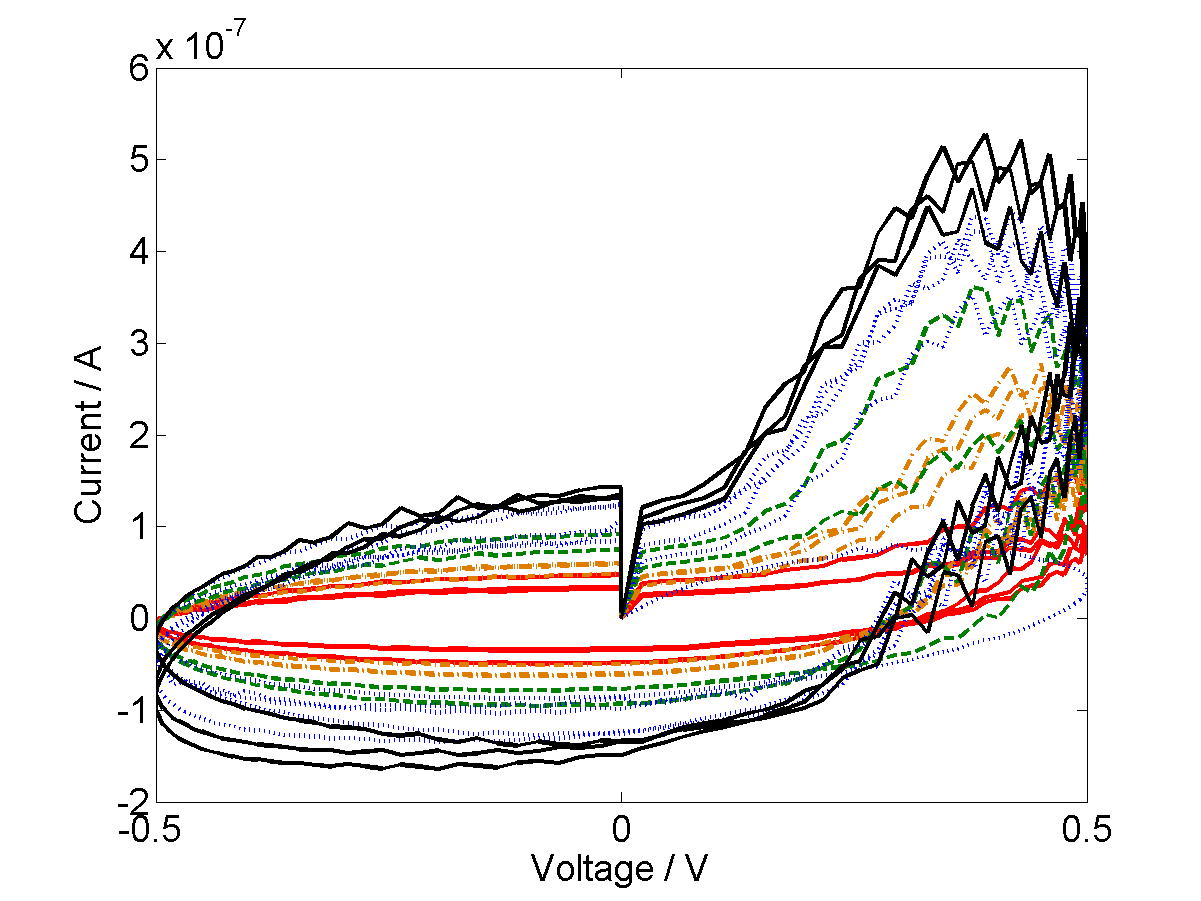}
	\caption{The effect of electrode size on memristor I-V curves. The larger the electrode, the bigger the I-V curve. Red = 1mm, Orange dot-dashes =2mm, blue dots = 3mm, green dashes = 4mm and black = 5mm.}
	\label{fig:AveCurves3}
\end{figure}

We shall start by looking at the $I-V$ curves. As can be seen in figure~\ref{fig:AveCurves3} the bigger the electrode, the bigger the $I-V$ curve. This demonstrates that a memristor's response under an electrical field is related to all three spatial dimensions of TiO$_2$. This result contradicts the assertion in model 1 that the memristance depends only on $D$ and not on the other spatial dimensions.

There is the peculiarity of the negative current seen at positive voltage and vice versa. We suspect that this is caused by the inertia of the moving oxygen ions (and work is underway to test this hypothesis). However, this effect has been seen before in~\cite{232} and as the authors of this paper used off-the-shelf electronic components to create a flux controlled memristor based on Chua's constitutive relation, it suggests that this effect is part of the memristive action rather than a corollary to it.  

The hysteresis increases with electrode size (see later for a more thorough discussion) but it does not increase equally across the devices, instead the top right quadrant of the curve ($W_1$) increases more. This assymmetry gives rise to a negative hysteresis as from the definition in equation~\ref{eq:Work} if $W_1 + W_4 > W_2 + W_3$ we have a negative sign of the hysteresis. Note that one of the 4mm devices is an outlier from the rest and its curve appears closer to the 1mm (red) and 2mm (orange) curves. When this device was examined by eye it was found to have a cracked electrode resulting in only a small part of the sol-gel covered by the electrode being electrically contacted.

\subsection{Testing the Bernoulli Rewrite of the Phenomenological Model}

\subsubsection{Test 1: Does $H_t$ and $H_b$ Give Approximate Hysteresis Values for the Sinusoidal Hysteresis?}

\begin{figure}[htbp!]
 \centering
 \includegraphics[width=5in]{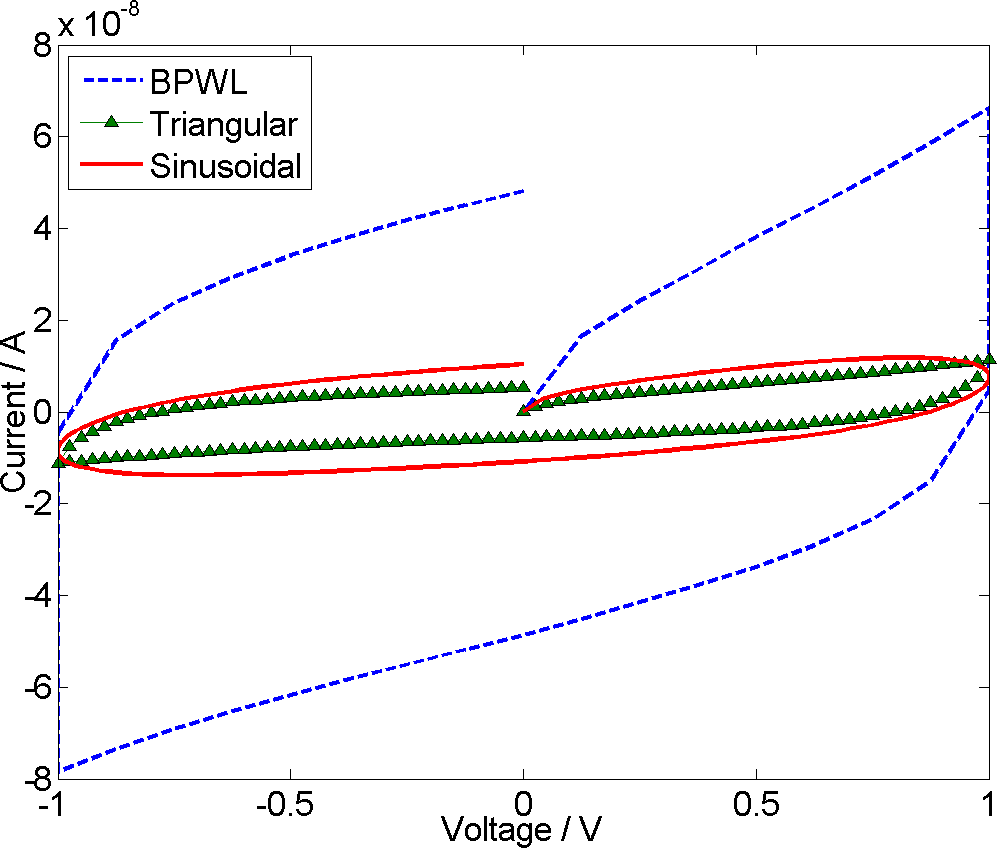}
 \caption{An example of I-V curves for the three different waveforms. The BPWL I-V curve is almost an order of magnitude bigger than the triangular or sinusoidal waveforms, however the triangular waveform is a good estimate of the lower limit of the sinusoidal hysteresis. $\tilde{\beta}$ was 0.9.}
 \label{fig:IVCurvesAllThree}
\end{figure}

As figure~\ref{fig:IVCurvesAllThree} shows, the $I-V$ curves for the BPWL waveform are much larger than the $I-V$ curve for the sinusoidal waveform, but the triangular waveform is slightly smaller than the sinusoidal waveform. At a single frequency, this relative order of hysteresis values would be expected, however, as the scaling of $\beta$ to make $\tilde{\beta}$ is supposed to account for differences in the waveform, it seems that this scaling doesn't work for the BPWL waveform.  


\subsubsection{Test 2: Is $\tilde{\beta}$ Related to $\bar{H}?$}

\paragraph{Across Devices of Different Sized Electrodes}

\begin{figure}[htbp!]
 \centering
 \includegraphics[bb=0 0 576 432,scale=0.5]{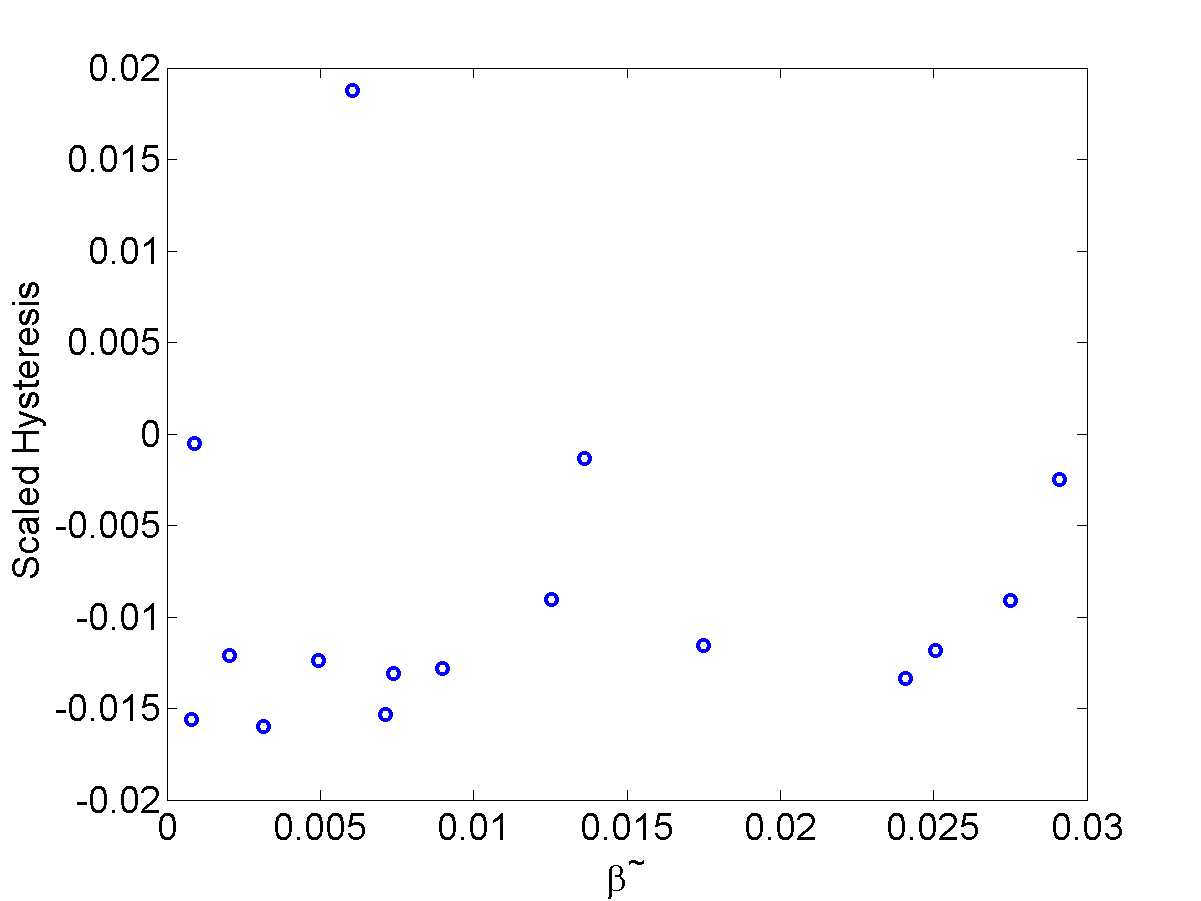}
 \caption{The relation between $\bar{H}$ and $\tilde{\beta}$ as measured at the same frequency across a set of devices with different electrode sizes. There is a no correlation.}
 \label{fig:HBarVersusBeta_1}
\end{figure}

\begin{figure}[htbp!]
 \centering
 \includegraphics[bb=0 0 576 432,scale=0.5]{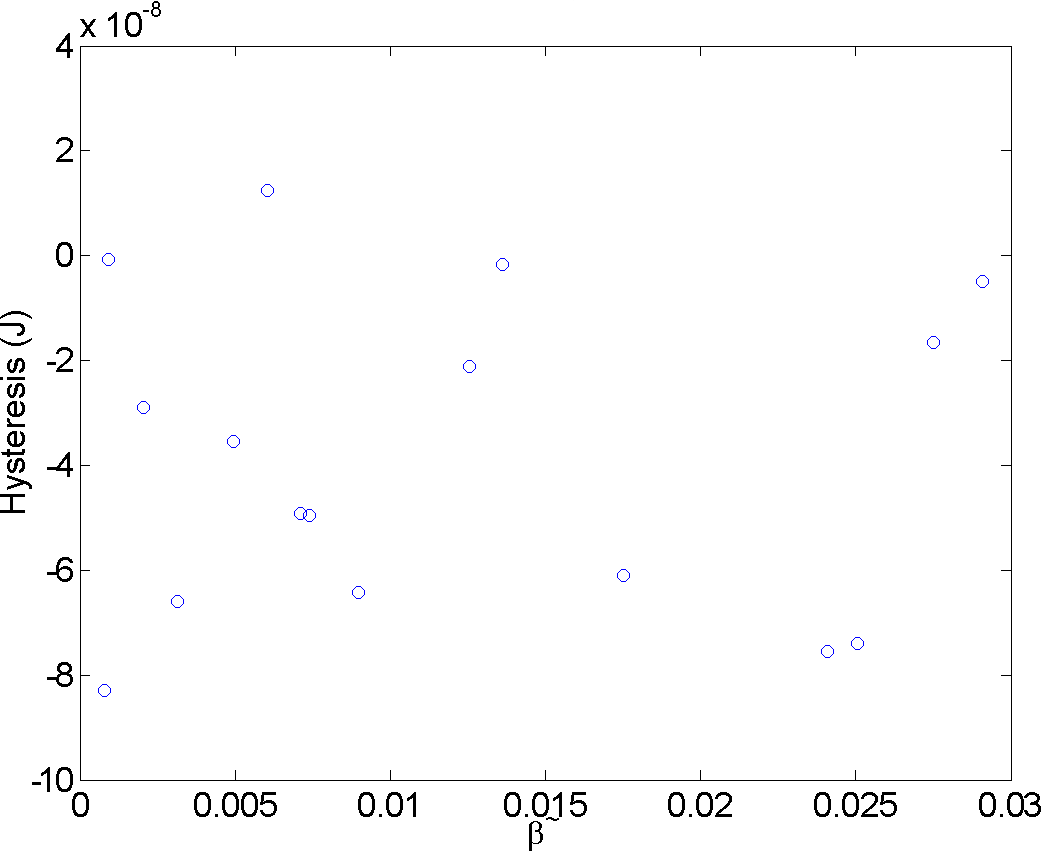}
 \caption{The relationship between the measured hysteresis and $\tilde{\beta}$ as measured for a group of devices with different electrode sizes.}
 \label{fig:HVsBeta}
\end{figure}

The measured hysteresis is negative and this is due to the increase in the size of the curve in the top branch which can be seen in figure~\ref{fig:AveCurves3}. However, if the `dimensionless lumped parameter' really does encapsulate the system's dynamics, it should be able to cope with this. No correlation between $\bar{H}$ and $\tilde{\beta}$ is seen.

According to model 2~\cite{224} there should be a monotonically increasing curve of $\bar{H}$ with $\tilde{\beta}$. Figure~\ref{fig:HBarVersusBeta_1} demonstrates that there is no correlation between the scaled hysteresis and $\tilde{\beta}$. If instead we look at the measured hysteresis, as shown in figure~\ref{fig:HVsBeta}, we also see no correlation. Note however that these results are measured over a rather small range of $\tilde{\beta}$ (and this range demonstrates how much $\tilde{\beta}$ changes with electrode width, i.e. not a great deal as these fabrication parameters are not included and only affect $\tilde{\beta}$ through their effect on $R_0$).

\paragraph{Across a Larger Range of $\tilde{\beta}$ on a Single Device}

\begin{figure}[htbp!]
 \centering
 \includegraphics[bb=0 0 576 432,scale=0.5]{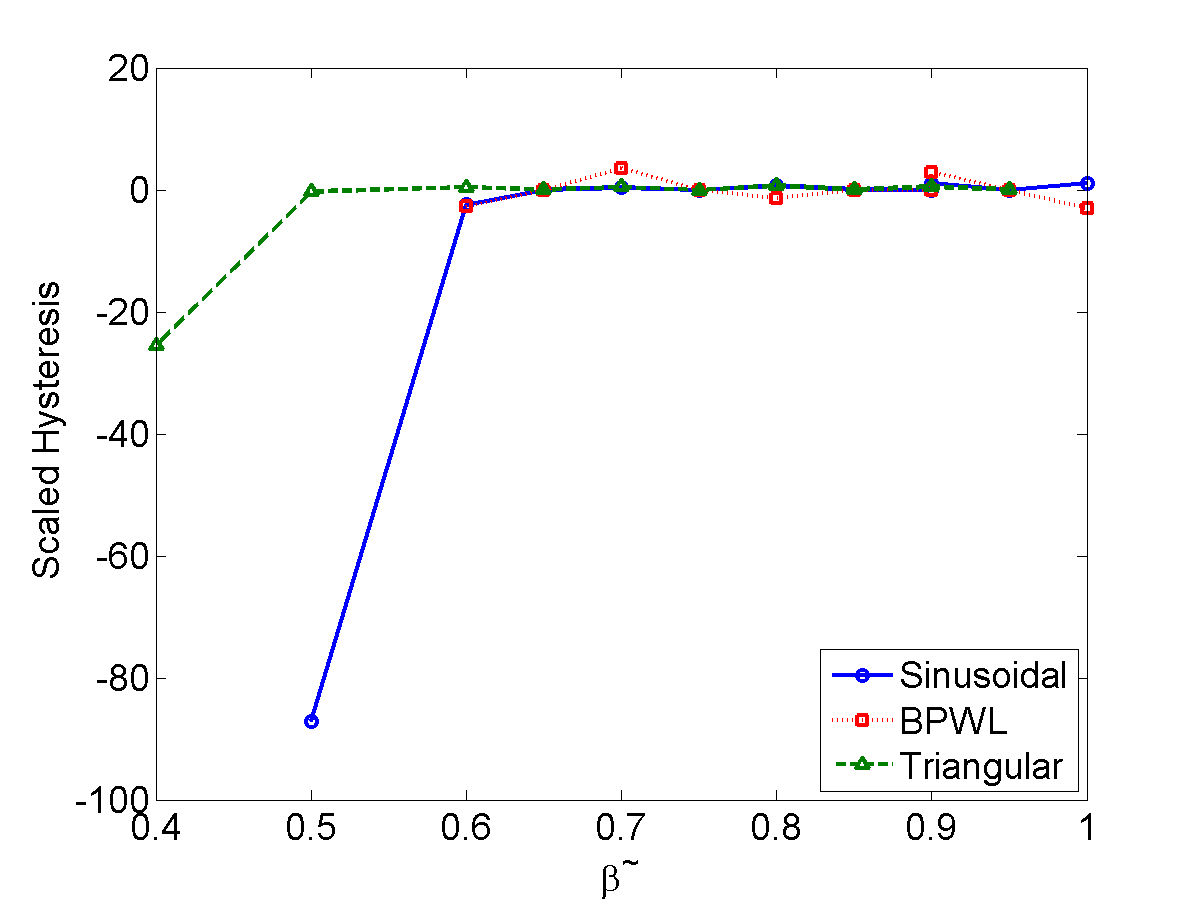}
 \caption{Scaled hysteresis versus the lumped dimensionless parameter. Lines are drawn to aid comparison. There is no monotonically increasing curve as predicted in~\cite{224}}
 \label{fig:HBarVersusBeta_2}
\end{figure}


The relationship between $\bar{H}$ and $\tilde{\beta}$ is shown in figure~\ref{fig:HBarVersusBeta_2} and this test concludes the experimentally testable claims of model 2. Over $0.6 < \tilde{\beta} < 1$ there is no relation with $\bar{H}$ moving around zero with $\tilde{\beta}$. It looks like the BPWL might be oscillating around zero, but examination of the two points at $\tilde{\beta}=0.9$ suggests that this `oscillation' is within the variance of repeated runs. As $\tilde{\beta}$ approaches the measurable frequency limit, the hysteresis increases to a large magnitude negative value and for this reason the effect of frequency on the hysteresis was examined.

\subsubsection{Test 3: Is $\bar{H}$ Related to the Measurement Frequency, $\omega_0$?}

\begin{figure}[htbp!]
 \centering
 \includegraphics[width=5in]{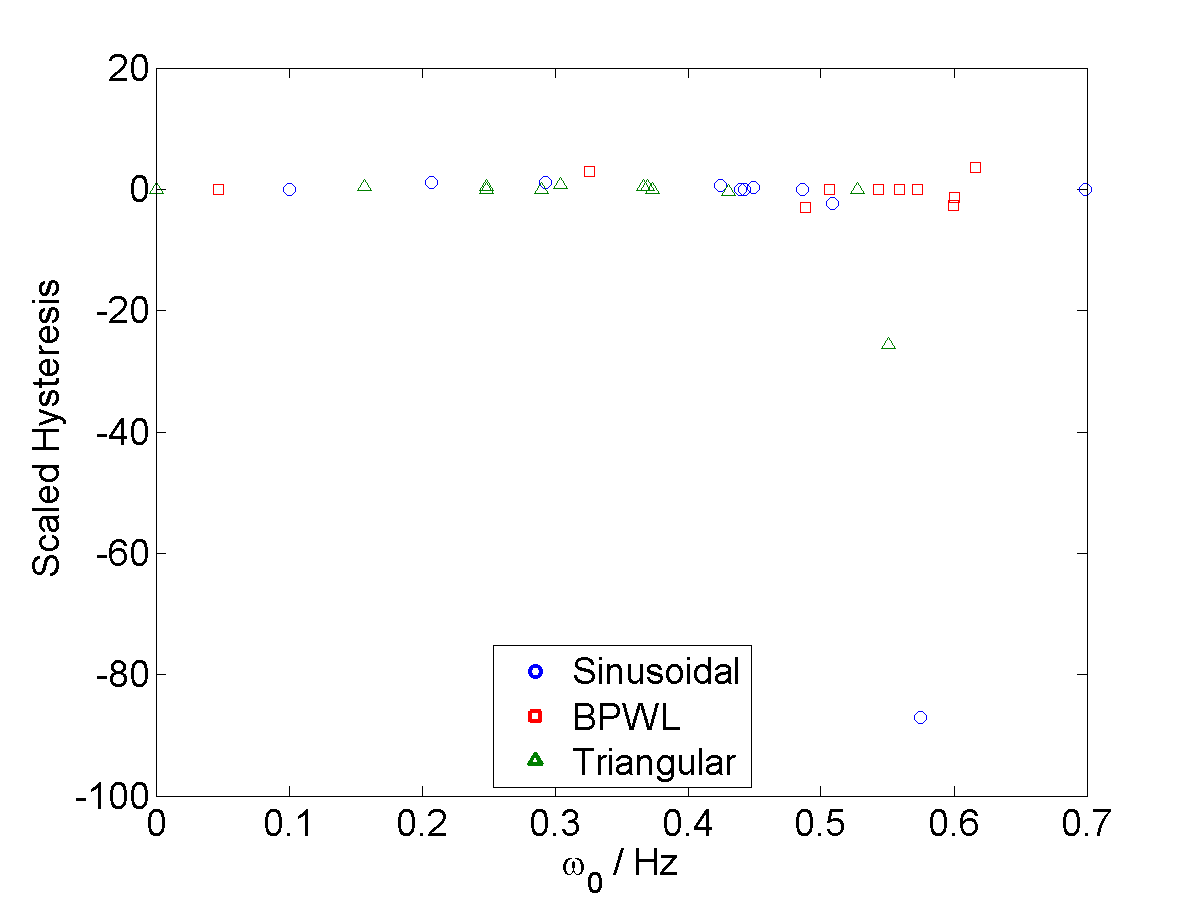}
 \caption{How the measurement frequency, $\omega_0$, affects the (experimentally measured) scaled hysteresis. There seems to be no correlation.}
 \label{fig:OmegaVsHbar}
\end{figure}

Model 2~\cite{224} is the first theory to tackle the question of how to predict hysteresis in memristors and it is also one of the few that explicitly includes the measurement frequency ($\omega_0$). Georgiou et al also make use of the concept of scaling the hysteresis relative to a linear resistor that doesn't change over the course of the experiment. Thus, we decided to see if the scaled hysteresis and the measurement frequency were related. As figure~\ref{fig:OmegaVsHbar} shows, it is not. 

\subsubsection{Test 4: Does the Theoretical $\bar{H}$ Allow the Prediction of the Measured $\bar{H}$?}

\begin{figure}[htbp!]
 \centering
 \includegraphics[width=5in]{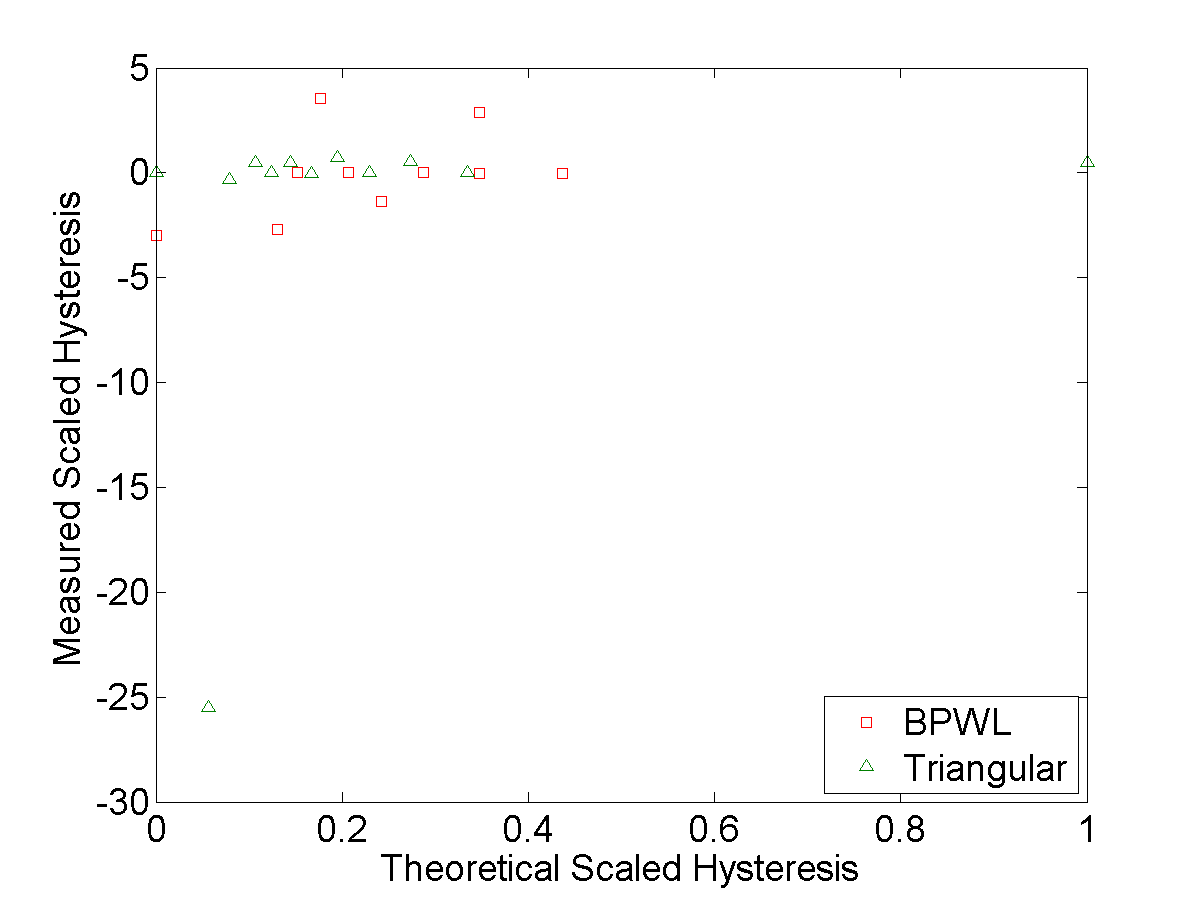}
 \caption{How the theoretical scaled hysteresis values relate to the experimentally measured scaled hysteresis. There is no useful relationship.}
 \label{fig:HbarTheoryVsHbarExp}
\end{figure}

It is possible that the open loop shape complicates hysteresis calculations or that one of the values (such as $\mu_v$ which is an approximate value for TiO$_2$ and could differ for different phases of the material) is incorrect, in which case the theoretical hysteresis as calculated by model 2 might only need tuning to be useful. For this reason we looked for a correlation between the theoretically calculated and experimentally measured values of $\bar{H}$. However, as figure~\ref{fig:HbarTheoryVsHbarExp} shows, there is no relation between the theoretical scaled hysteresis and the experimentally measured hysteresis. 

\clearpage

\subsubsection{Summary of Model 2 Tests}

These experiments convincingly demonstrate that model 2 does not work for these devices. The `dimensionless lumped parameter' in both in scaled and unscaled forms does not show any correlation to the hysteresis in either of its scaled or unscaled forms. This has been tested across a set of devices over a small range of $\tilde{\beta}$ and over a large range of $\tilde{\beta}$ with one device. Furthermore, the scaled hysteresis is not related to the measured hysteresis which suggests that the theory is not useful for hysteresis prediction. 

\subsection{Testing the Memory-Conservation Theory of Memristance}

As shown in figure~\ref{fig:AveCurves3} the size of the $I-V$ curve increases with the size of the top electrode. 

\subsubsection{How the Memory and Conservation Functions Change with Electrode Size}

\begin{figure}[htbp!]
\centering
 \includegraphics[width=5in]{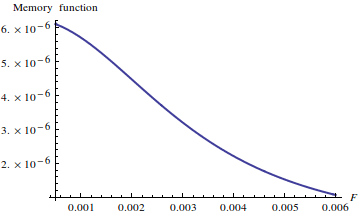}
 \caption{The effect of changing $F$ on the memory function. Here, $mu_v$ is set to 1 so the resistance is in reduced units and relevant to $mu_v$. The shape is qualitatively similar to the experimental results.}
 \label{fig:MemFuncRU}
\end{figure}

\begin{figure}[htbp!]
 \centering
 \includegraphics[width=5in]{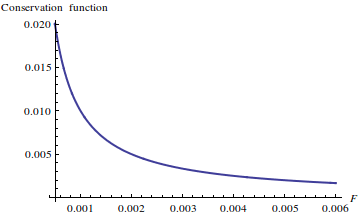}
 \caption{The prediced effect of electrode size on the conservation function. $rho_{off}$ was set to 1 so the resistance is in reduced units. The shape is qualitatively similar to the experimental results.}
 \label{fig:ConFuncRU}
\end{figure}

It has been stated above that because the theory includes the other electrode dimensions ($E$ and $F$) then changing one or both of these values should have an effect, but what effect? A qualitative answer is given in figures~\ref{fig:MemFuncRU} and ~\ref{fig:ConFuncRU}. For these plots the memory function was evaluated with $L=0.5/D$, $\mu_v = 1$, $D$, $E$ and $F$ set to the values for our devices, and the conservation function was evaluated with $\rho_{\mathrm{off}}=1$. For the memory function $w$ was set to $D$, which will give answers in the limit of maximum possible resistance (i.e. its upper limit) for that function. For the conservation function, $w$ was set to 0, which will give answers in the limit of maximum possible resistance (i.e. the upper limit) for that function. As the total resistance is the sum of these two terms, these evaluations correspond to the total resistance when the device is fully switched on (memory function, with $w \rightarrow D$) and fully switched off (conservation function, with $w \rightarrow 0$). The plots are in reduced units relative to the device variables $\mu_v$ and $\rho_{\mathrm{off}}$ and because $c_M$ has not been included these plots are not directly comparable to each other. Nonetheless, these plots serve to tell us that both the memory and conservation functions (which are the resistances of the TiO$_{(2-x)}$ and TiO$_2$ parts of the device respectively) decrease with size, leading to an expectation that devices would shift to a higher current on and off state with increasing electrode size. This shows that, as qualitatively predicted from the theory, the devices will have both higher current on and off states if fabricated with larger electrodes. 

\subsubsection{Testing the Conservation and Memory Functions}

As explained above, the maximum values of the conservation and memory functions correspond to the maximum, $\Roff$, and minimum, $\Ron$, resistances respectively. If we measured the device at its fundamental frequency (namely that which caused the greatest hysteresis) we would fully switch the device from the off state, to the on state and back again in a single $I-V$ cycle. Our devices may not fully switch over the course of a single measurement cycle at this frequency, but we can assume that we are in the limit where the measured $\Ron$ tends to that which would be measured at the fundamental frequency, and similarly for $\Roff$. This is equivalent to comparing the measured device at frequency $\omega$ to a thinner device with a fundamental frequency of $\omega$ (i.e. thinner device would have the same physical limits of $w(t)$ as those travelled over the experiment run at $\omega$) or a device modelled with a strong window function. (Remember that $\omega$ is omega, a frequency and $w$ is the distance the 
boundary between doped and undoped TiO$_2$ travels).

\begin{figure}[htbp!]
 \centering
 \includegraphics[width=5in]{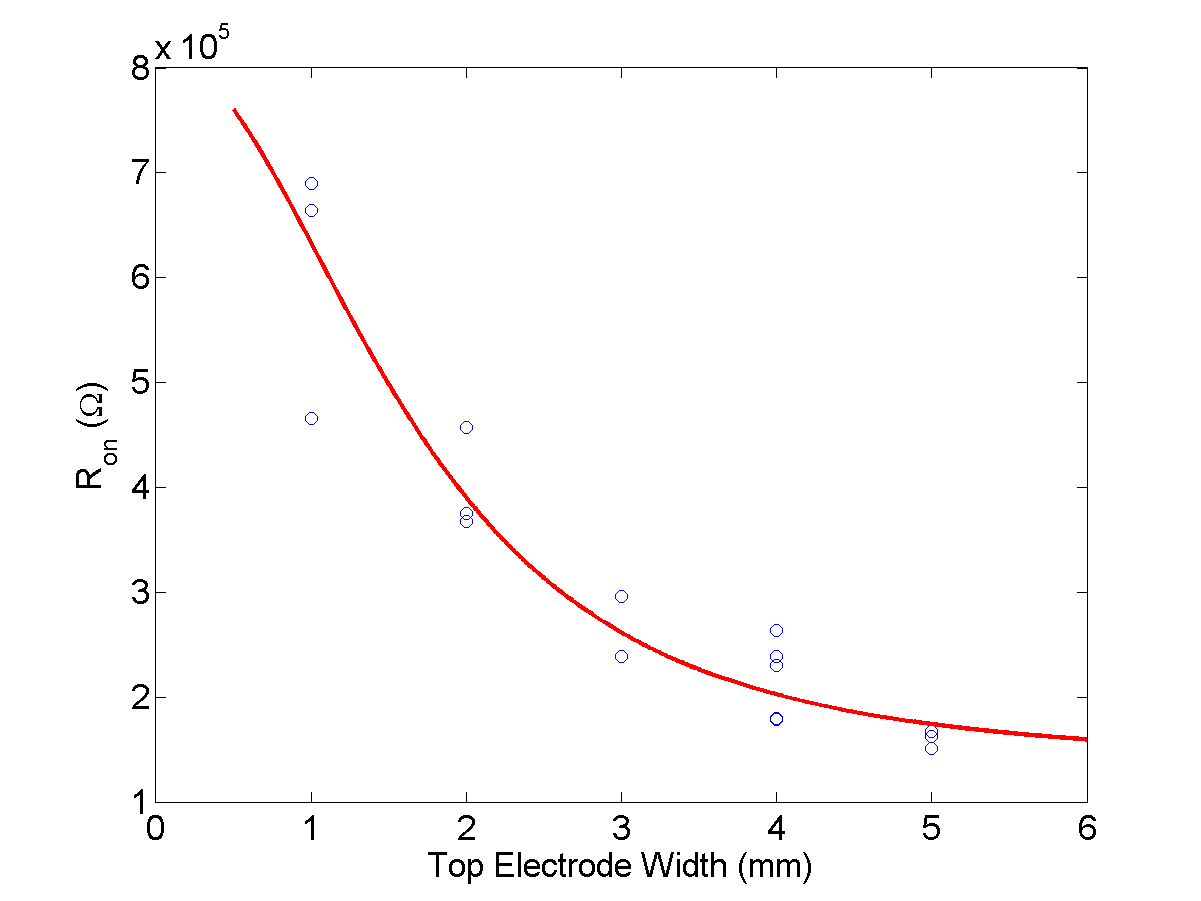}
 \caption{Experimental values for $\Ron$ fit by the memory function, $M_e$.}
 \label{fig:MemFit}
\end{figure}

The fit of equation~\ref{eq:Mem} for the memory function $M_e$ to the data for $\Ron$ is plotted in figure~\ref{fig:MemFit}. We have used one fitting parameter, as included in the theory, and its value is $C_M=2.09 \times 10^{25}$. Furthermore, that parameter, $C_M$, is a measure of the different resistances of the material to oxygen ions and electrons and the theoretical basis for it is under investigation. As can be seen in figure~\ref{fig:MemFit} once $C_M$ is determined, the memory function describes its value and relation to the electrode width. Even the slightly odd shape of the experimental data correlation is well described by the theory and this shape is present in the unfitted, reduced unit version in figure~\ref{fig:MemFuncRU}.

\begin{figure}[htbp!]
 \centering
 \includegraphics[width=5in]{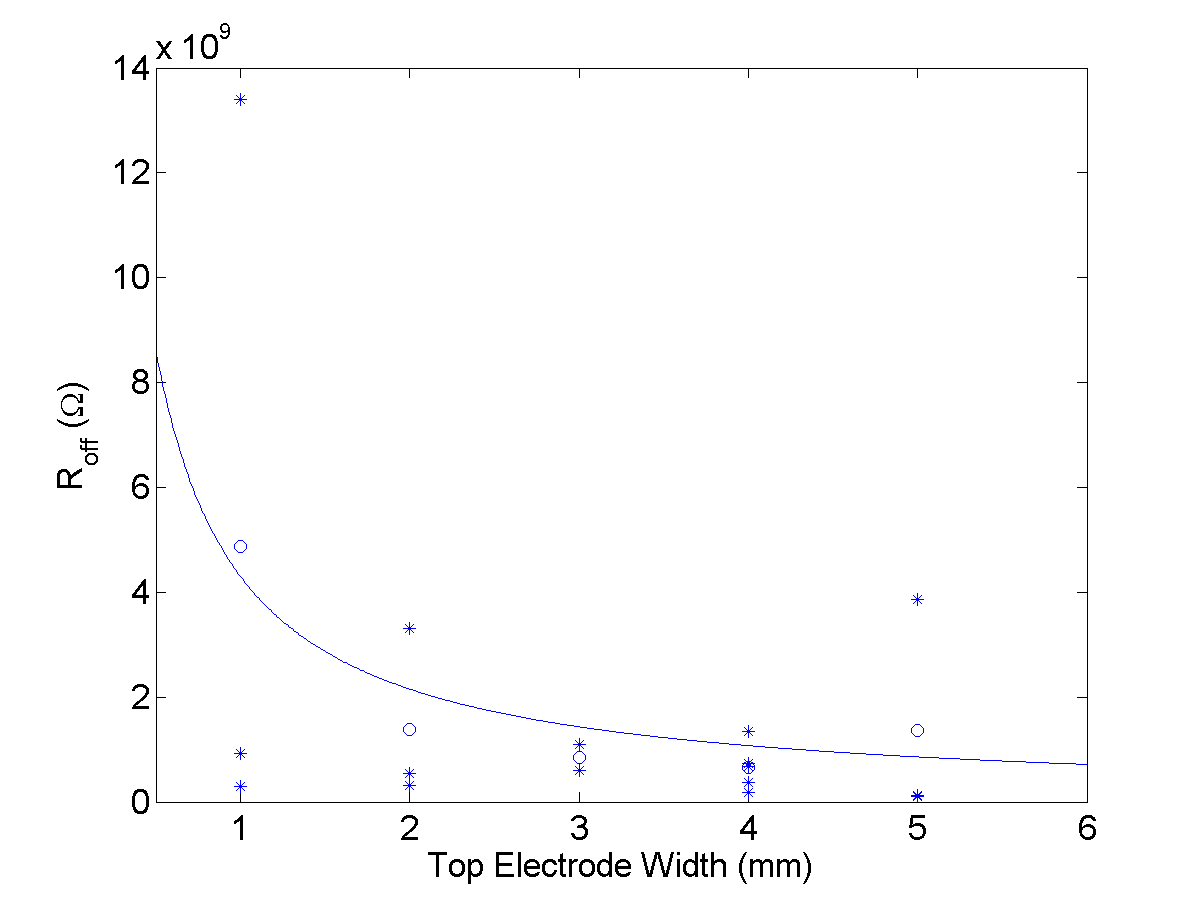}
 \caption{The conservation function fit to $\Roff$. As there was a greater variance in $\Roff$ values, the mean $\Roff$ are shown to indicate the goodness of fit for each electrode width value (unfilled circles) along with the $\Roff$ values (dots).}
 \label{fig:RoffConFit}
\end{figure}

The fit of equation~\ref{Con} for the conservation function, $\Rcon$ to the data for $\Roff$ is plotted in figure~\ref{fig:RoffConFit}. There is greater variance in the $\Roff$ values at each electrode width and thus the mean $\Roff$ values have been plotted as well to allow an easier determination of the fit by eye. The conservation function fits the data. To do this fit the resistivity of the OFF semiconductor material was used as a fitting parameter which is reasonable as we do not know exactly which phase of titanium dioxide the gel is in. From our fit, we get the resistivity as $\rho_{\mathrm{off}} = 6.82 \times 10^{10} \Omega m$, which is a reasonable value given that, for example, the resistivity of anatase can be anywhere in the range of $10^{4}$ to $10^{12} \Omega m$.


\subsubsection{Summary of memory-conservation tests}

The memory-conservation theory suggests that the ON and OFF resistance states should decrease with an increased electrode size. This is exactly what is seen in our data. The memory function fits the ON state resistance very well with only one fitting parameter. The conservation function fits the OFF state data well with only one fitting parameter, which is the resistivity and which comes out at a sensible value for the resistivity of titanium dioxide thin film.

\subsection{Which Device Properties Cause the change in Hysteresis?}

\begin{figure}[htbp!]
 \centering
 \includegraphics[width=5in]{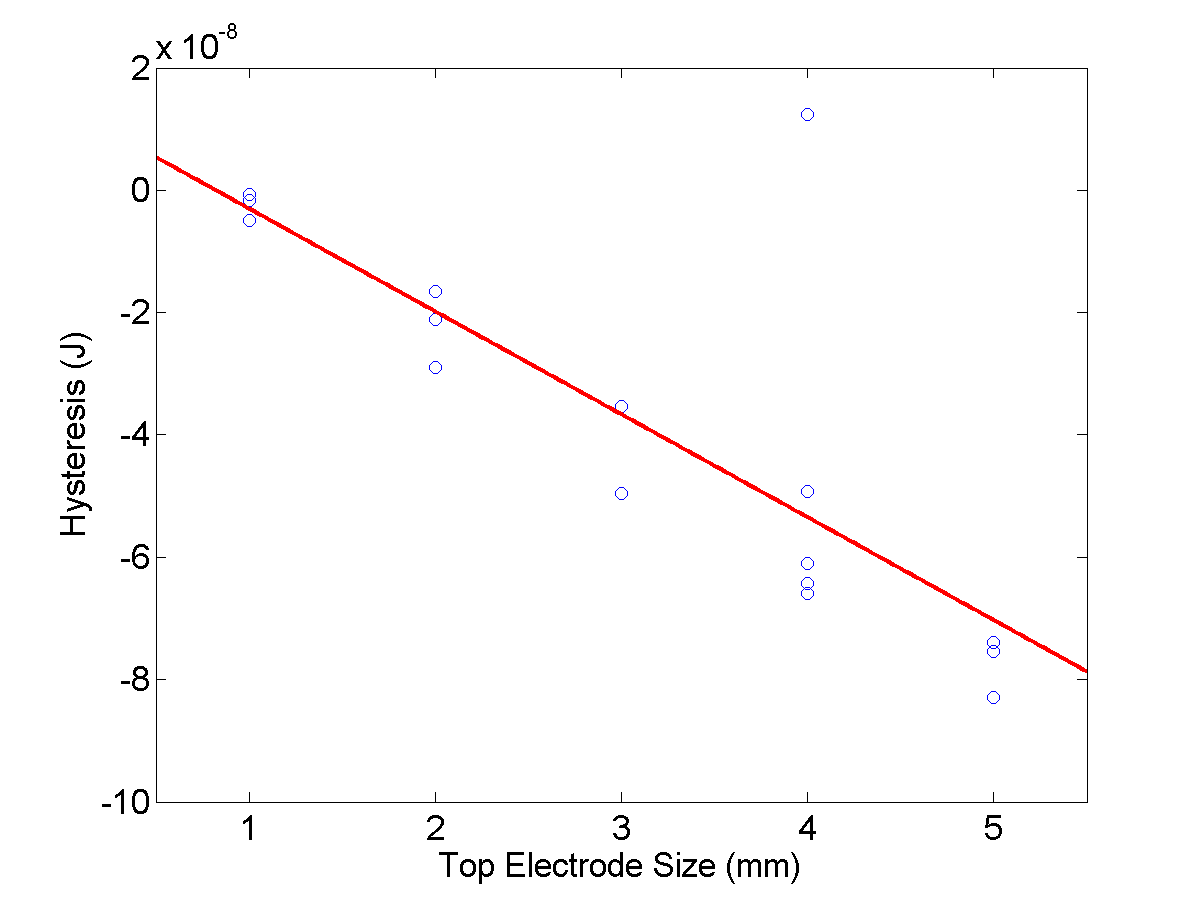}
 \caption{The magnitude hysteresis increases with electrode size.}
 \label{fig:HVsF}
\end{figure}

Using the data from different sized electrodes we are now in a position to examine which device properties are actually related to the hysteresis. As figure~\ref{fig:HVsF} shows, the magnitude of the hysteresis increases with electrode size (Note that the outlier seen in figure~\ref{fig:AveCurves3} is the only device with positive hysteresis). Thus, if we want a device with larger hysteresis, we should increase the size of the electrodes; the fit in figure~\ref{fig:HVsF}, which is given by $H=m_F F + c_F$ where $m_F = -1.68 \times 10^{-8}$ and $c_F = 1.39 \times 10^{-8} J$ ($|r^{2}|=7.18 \times 10^{-8}$), gives us the quantitative relationship we should use. 

\begin{figure}[htbp!]
 \centering
 \includegraphics[width=5in]{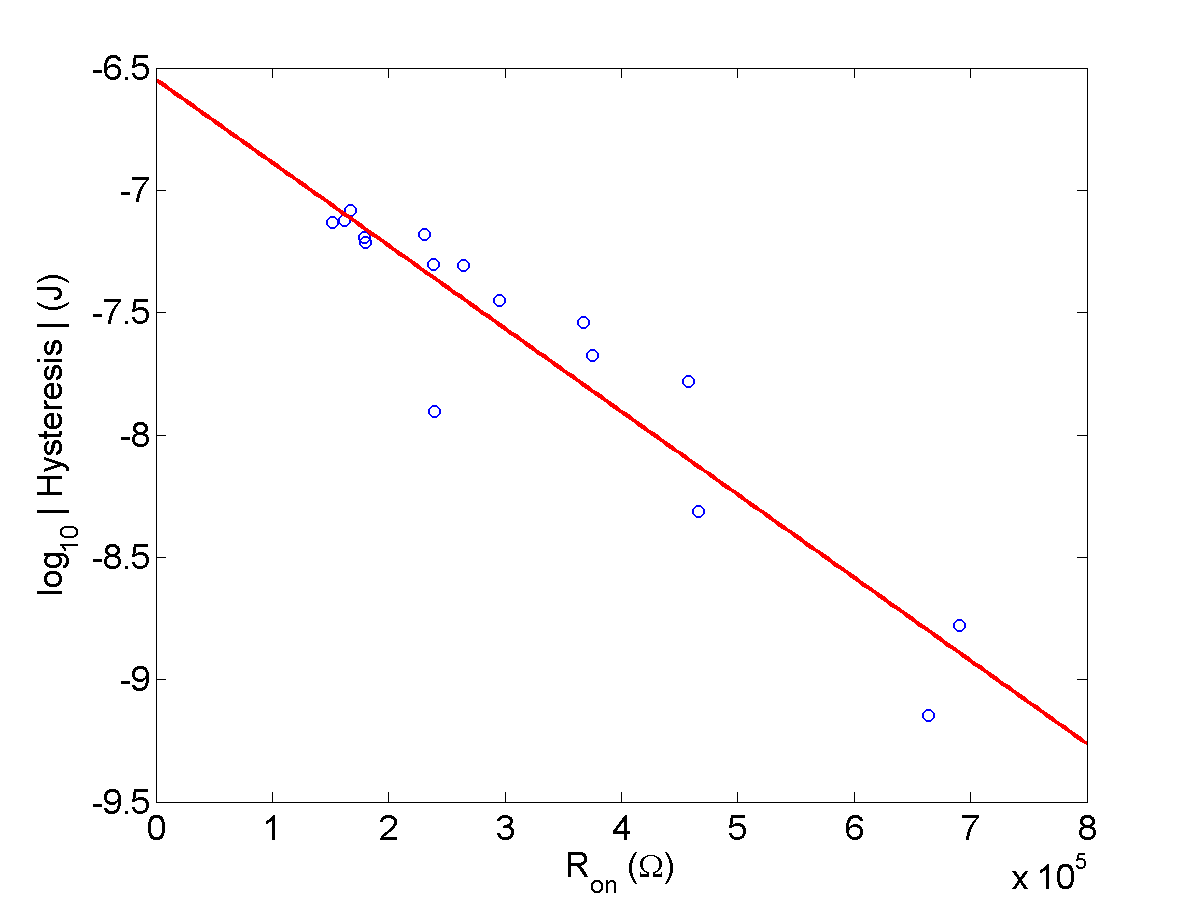}
 \caption{The logarithm of the hysteresis is related to the measured $\Ron$ value.}
 \label{fig:RonVsH}
\end{figure}

We can also now comment on how larger electrodes causes a larger hysteresis. Given that $\Ron$ is highly correlated with $F$ (and $\Roff$ is less so), we might expect that decreasing $\Ron$ would lead to a larger hysteresis and figure~\ref{fig:RonVsH} shows this to be the case. The ON state resistance is actually proportional to the logarithm of the hysteresis magnitude as given by $\log_{10} | H | = m_{R} \Ron + c_R$, where $m_R = -3.39 \times 10^{-6}$, $c_R = -6.55$ ($|r^2|=0.845$). This is a better fit than for the electrode size, showing that measurement of the actual $\Ron$ is a better predictor for the hysteresis than the electrode size, although, the electrode size the property to control for in fabrication. Note that by taking the magnitude of the hysteresis and the measured $\Ron$ the outlier has moved closer to the line, this point has the hysteresis and $\Ron$ of a device with a top electrode width of $\sim 1.6mm$ which roughly agrees with the observation of the width of the cracked electrode.
 
Whether the electrode size causes the change in hysteresis size directly or via changing $\Ron$ is not known, but there are a few facts that suggest the latter. $\Ron$ is related to the electrode size by $\Ron = m_g F + c_g$ (graph not shown) where $m_g = -1.08 \times 10^5$ and $c_g = 6.58 \times 10^5 \Omega$ ($|r^2|=2.68 \times 10^5$). The hysteresis is a measure of the interaction of two sets of parameters: $\{ \Roff , \Ron \}$ and $\{ \omega_0 , \mu_v \}$. The limits of the loop are prescribed by the measured maximum and minimum resistances at that frequency, i.e. $R(\mathrm{max}) \mid_{w(\mathrm{min})}$ and $R(\mathrm{min}) \mid_{w(\mathrm{max})}$. The maximum and minimum resistances for the fully switched device are $\Roff \mid_{w \rightarrow 0}$ and $\Ron \mid_{w \rightarrow D}$ respectively. The interaction between $\omega_0$ and $\mu_v$ affect the amount that $w$ moves and thus the value of $R(\mathrm{max})$ and $R(\mathrm{min})$ compared to these limits. Obviously the ionic mobility gives rise to a characteristic timescale for the device which gives the resonant frequency when transformed into the reciprocal time domain. Note that we found no correlation between the ratio $\frac{\Ron}{\Roff}$ and $H$, neither is there a correlation between $R_0$ and this ratio. 

As the values of $F$ were taken to be what the top electrode should be, and was not measured, it is not exactly known whether the actual values of $\Ron$ only depend on the sputtered electrode width or the quality of the sol-gel layer underneath it. We expect that any fabrication parameter that changes the value of $\Ron$ will affect the hysteresis.

\subsubsection{Simulating I-V Curves to get Hysteresis}

The memory-conservation model can be used to predict the hysteresis. In this section we will work out some of the relevant values from the experiment to put into the model to calculate the hysteresis by simulating the experiment. We have done further work on extending the memory-conservation model to more accurately model our specific systems~\cite{254}, but here we will stick to the basic memory-conservation model as described in~\cite{F1} to demonstrate that even without that specificity we can get a good approximation of real world devices. 

First we shall consider the value of the ion mobility. The value that was used in the fitting was 1$\times10^{-10}$cm$^2$V$^{-1}$s$^{-1}$ as taken from~\cite{15}. However, we are now in a position to calculate an approximate value from our data. 

Consider a frequency $\omega_f$ which is sufficient to fully charge and discharge the memristor on a single cycle (this is the frequency that would give the maximum possible hysteresis). We don't know what this frequency is, however earlier we approximated the measured $R$(min) as $\Ron$ and $R$(max) as $\Roff$, which is implicitly assuming that our operating frequency, $\omega$ was equal to $\omega_f$. Keeping to this approximation, we can calculate the approximate drift velocity, $v_d$, from the period from $v_d=\frac{(D-0)}{\frac{1}{2}T}$, where $T$ is the period ($\approx$9 seconds) and we've taken 0 and $D$ as the fully switched and un-switched limits of $w$ (which is optimistic for a real device). Here we have also assumed that the speed increases as an sigmoidal-curve and taken the median value, in actual fact there is lag in the current response so the curve would be skewed. 

Acknowledging these approximations, we get the value of $\mu_v \approx 9.777\times10^{-16}m^{2}V^{-1}s^{-1}$ which is 0.0098 times the approximated value taken from the literature~\cite{15} (note that the literature value is assumed, not measured, and is for an atomically deposited thin film rather than the amorphous gel layer like we have in these devices). 

To simulate our memristors we used the $C_m$ and $\rho_{\mathrm{off}}$ values found earlier with the following change for the memory function. To account for the change in $\mu_v$ $M_e$ was taken as $0.0098*M_e$ as $\mu_v$ is multiplied by the whole function and $C_m$ was fitted with that value of $\mu_v$. With these settings the device is modelled with 0nm$ < w < 36$nm.

\begin{figure}[htbp!]
 \centering
 \includegraphics[width=5in]{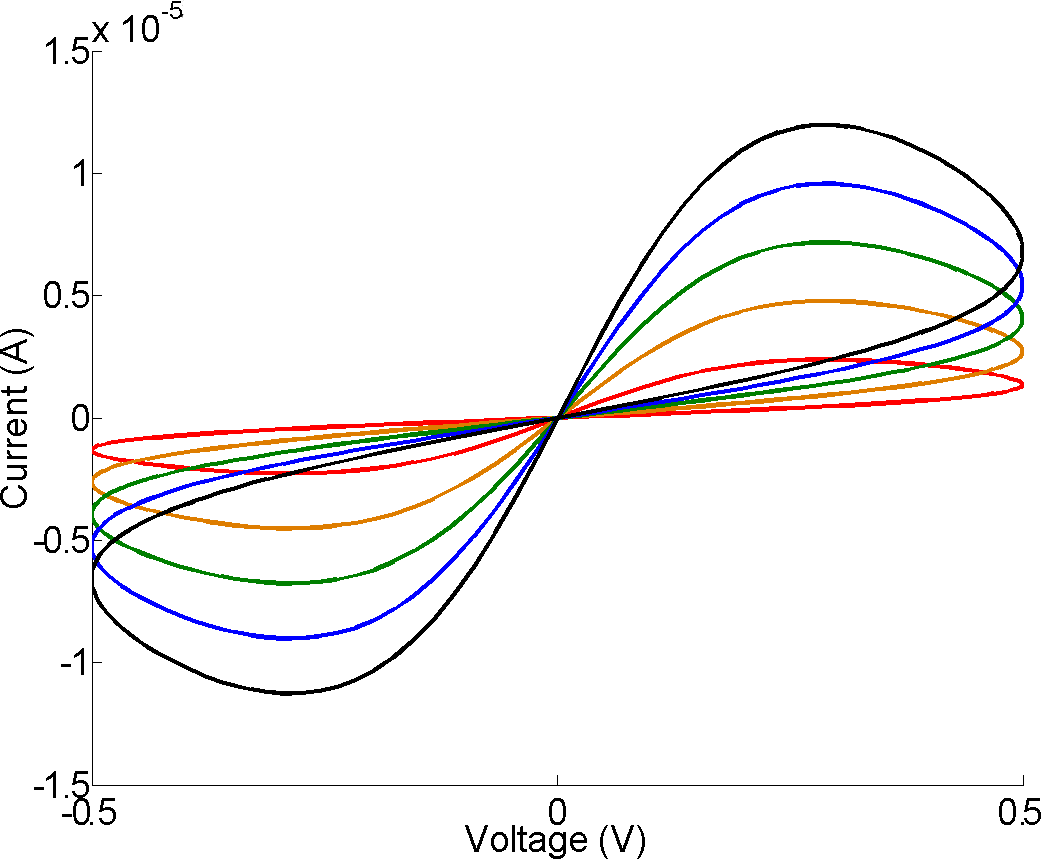}
\label{fig:SimIV}
\caption{Simulated I-V curves for the different sized devices. The theory correctly predicts the direction of the effect of different electrode sizes. Red: 1mm; Orange: 2mm; Green: 3mm; Blue: 4mm; Black: 5mm}
\end{figure}

The assumptions we made to do the fitting, namely that the conservation function contributed nothing to the $\Ron$ and that the memory function contributed nothing to the $\Roff$ values, means that the memristance due to the memory function is several orders of magnitude lower than the conservation function. This is an approximation, however the combined functions are on the order of 10$^{-5}$A, see figure~\ref{fig:SimIV}, which is two orders of magnitude out from the experimental values which are 10$^{-7}$, see figure~\ref{fig:AveCurves3}.

Although it seems that the lumped dimensionless parameter $\tilde{\beta}$ is not useful in predicting the hysteresis, it might be that Georgiou et al were correct to expand the range of device parameters/measurables included in the model, specifically, by including $R_0$. The memory-conservation model only covers the sol-gel layer of the memristor, so we can improve the model by including a contact resistance. Thus we decided to take $R_0$ as a measure of the contact resistance which includes the resistance of the electrodes, wires and contacts (and also the starting resistance of the device). This is slightly different to Georgiou et al's formulation where they use $R_0$ to indicate which state they are starting in, we instead assume that we are starting the experiment in the ON resistance state (as this is experimentally what we did) and using $R_0$ as the measure of the unswitchable resistance. 

\begin{figure}[htbp!]
 \centering
 \includegraphics[width=5in]{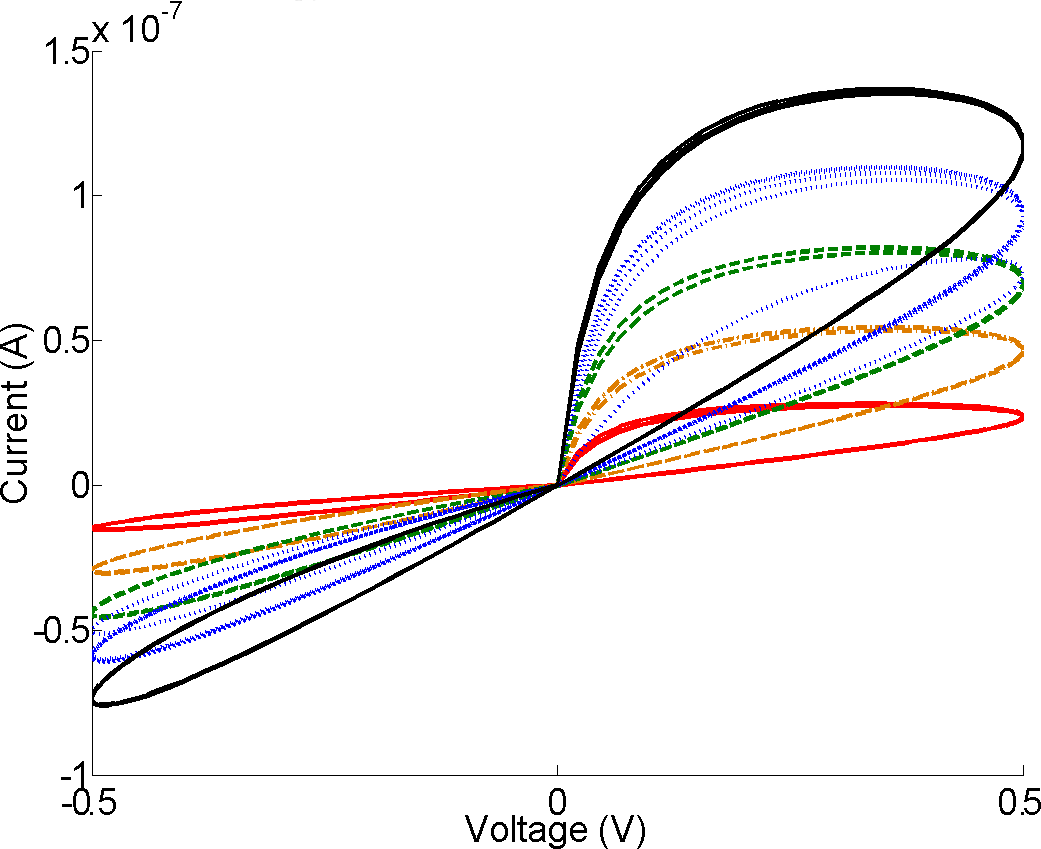}
\label{fig:SimIVWithRc}
\caption{Simulated I-V curves with $R_0$ included as a measure of contact resistance. The model is much improved with the simulated current now within the correct order of magnitude as the experiment. Red: 1mm; Orange dot-dashes: 2mm; Green dashes: 3mm; Blue dots: 4mm; Black: 5mm}
\end{figure}

If we take the measured $R_0$ as being related to the contact resistance, i.e. that part of the device which is not switched at this frequency, and add that as a constant, we get the $I-V$ curves shown in figure~\ref{fig:SimIVwithRc}. This is exciting for such a basic model as it is within the same order of magnitude as the real results. Even better it explains the mystery of where the larger positive quadrant lobe comes from (this is a common occurrence in other published real world devices). The inclusion of $R_0$ has also separated the devices with the same electrode size from each other, including the real world variance between devices which is visible in the hysteresis loops. The fact that the shape is different is a problem we will discuss in a forthcoming paper. 

This result demonstrates that the skewed I-V curves with unusually large positive quadrant lobes as seen in our results~\cite{M1} and other memristors may be due to the contact resistance found in experimental systems but not `vanilla' memristor models.

\begin{figure}[htbp!]
 \centering
 \includegraphics[width=5in]{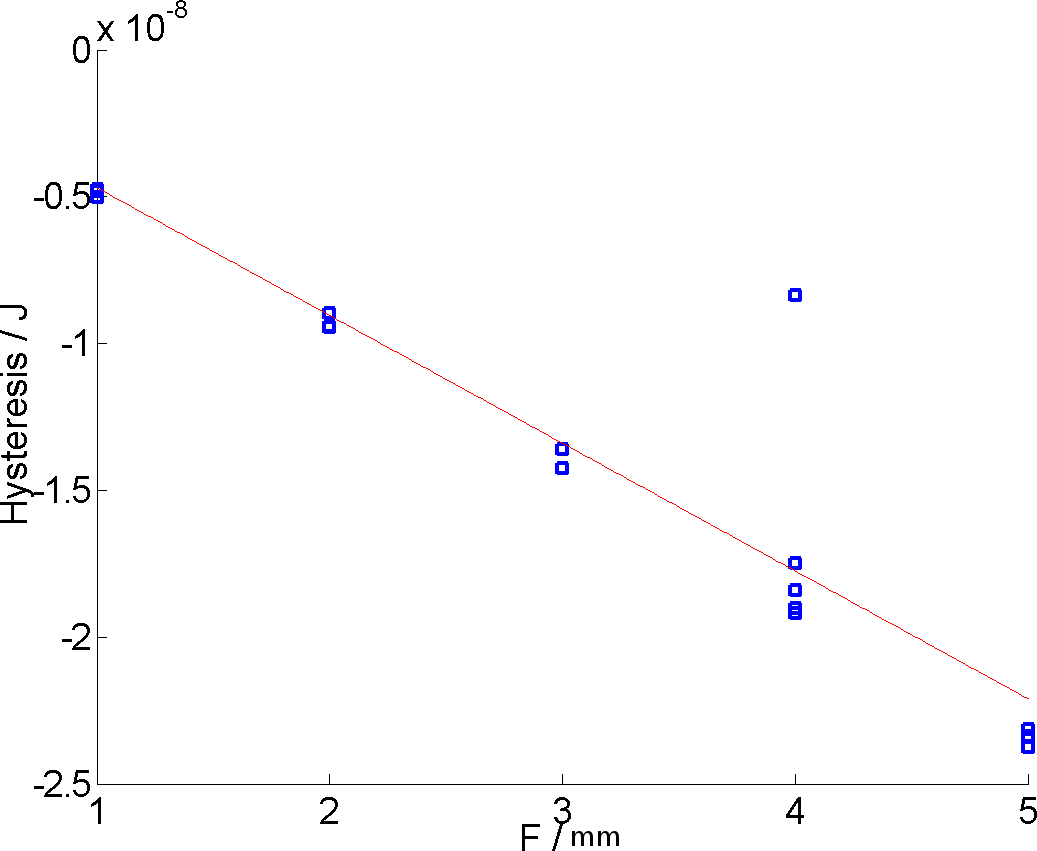}
\label{fig:SimIVWithRc}
\caption{Simulated hysteresis values against the electrode size. The simulation correctly predicts the experimentally measured effect of the electrode size.}
\end{figure}

\begin{figure}[htbp!]
 \centering
 \includegraphics[width=5in]{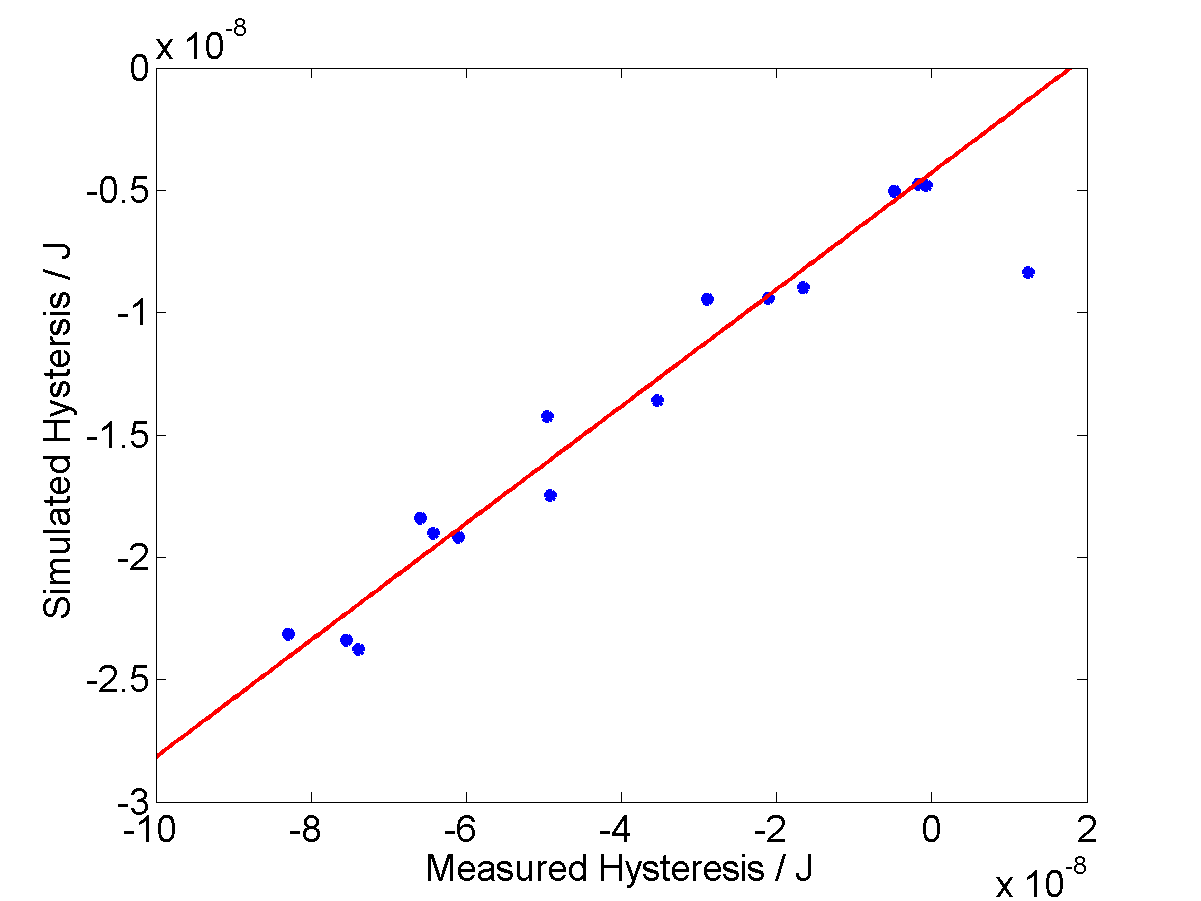}
\label{fig:SimIVWithRc}
\caption{The simulated hysteresis values plotted against the experimental hysteresis value. There is a straight-line relation which is very good especially if we discount the outlier in the data.}
\end{figure}

How good is the memory-conservation theory at predicting the value of the hysteresis? Firstly it is within the same order of magnitude and also predicts the negative sign due to positive loop asymmetry. The fitted equation for how the hysteresis should change with electrode size is $H = m_s F + c_s$ where $m_s = -4.351 \times 10^{-9}$ and $c_s = -3.4204 \times 10^{-10} J$ ($|r^{2}| = 9.9466 \times 10^{-9}$) and it even predicts that there should be an outlier as the memristor with the cracked electrode had an erroneously high $R_0$. 

Even better, the simulated values, $H_{\mathrm{sim}}$ are off by a specifiable amount, see figure~\ref{fig:SimIVWithRc}. Excluding the outlier, the fitted equation is $H_{\mathrm{sim}} = m H + c$, where $m=0.23886$ and $c=-4.255 \times 10^{-9}$J ($|r^2|=7.52 \times 10^{-9}$). Thus, if we know $\mu_v$ (approximatable or measurable), the starting resistance of a device (which can be approximated from the electrode size and sol gel properties, i.e. $F$ and $\mu_v$ or measured) and the frequency, we can simulate and calculate the hysteresis and then apply this conversion equation to predict the real world value to within 5.5nJ (10\% of the hysteresis value on average). As Georgiou et al state that this is a useful thing for engineers to know to design memristor circuits, we suggest such circuit designers use the method outlined in this section. 

\subsubsection{Summary of Hysteresis Simulations}

The size of the hysteresis is correlated with $\Ron$ and, as that is related to the size of the electrodes, the hysteresis value can be tuned by altering the size of the electrodes. Simulations of the devices using the memory-conservation model, the calculated fitting parameters and the device fabrication parameters allows us to predict the hysteresis. Including the contact resistance ($R_0$) as well gives us values within one order of magnitude of the actual values. The predicted values are related to the measured values linearly, meaning that, after running tests such as these we to find these fitting values between simulation and experiment, we can then use the memory-conservation simulations to predict the actual device hysteresis values.    

\section{Conclusion}

The width of the memristor electrodes does have an effect of the memristance of the device. This width affects $\Ron$ especially, but also the starting resistance and off state resistance, which in turn affects the size of the hysteresis with larger devices having a larger hysteresis. This hysteresis tends to be asymmetrical around zero (in these devices) leading to a negative value for the hysteresis. 

The lumped dimensionless parameter and the Bernoulli rewrite of Strukov's equations has been shown to not work at predicting the hysteresis (in either scaled or non-scaled format). This could be due to two causes. It could be that the phenomenological model does not work for these devices and that a Bernoulli rewrite of the memory-conservation model could encapsulate useful information into a single parameter. From this data there is no way of telling if this would work, however we don't think so. The memristor is a nonlinear circuit element and rewriting the equations into a linearisable form is, in our opinion, removing too many of the behavioural aspects to make the model useful for a real world situation, in essence, we postulate that the Bernoulli rewrite simplifies the system too much. The memory-conservation model was derived to keep in aspects of the model that were `abstracted out' in the phenomenological model and thus we suspect that lumping everything into a single parameter maybe an oversimplification. However, it could well be that the memory conservation model can be abstracted this way, so that only one parameter is needed to describe where the memristor system is in phase space. If this is the case, we would expect $\beta$ to include: $D$, $w$, $E$, $F$, $\mu_v$, $V$, $\omega_0$, $R_0$, $\rho_{\mathrm{off}}$, $\rho_{\mathrm{on}}$ or parameters derived from them. It would be interesting to know if applying the Bernoulli linearisation to the memory-conservation model would yield useful results. 

The other reason why the Bernoulli equation rewrite might not work, as tested here, could be because the phenomenological model is incorrect. This is a model based on a 1-D description of the system. However, it does not expect there to be any effect of the electrode widths, which we have demonstrated has an effect. Furthermore, the derivation of model 1 has been shown to have a fatal algebraic flaw that leaves model 1's expression of memristance unsupported~\cite{F1}.

These results have provided an experimental verification of the memory-conservation model. The memory-conservation model has been shown to: predict an increase in current response and hysteresis size with increasing electrode size, which was experimentally measured; fit (via the conservation function) the values of $\Roff$; fit (via the memory function) the values of $\Ron$; estimate (via simulation using experimentally measured values) the hysteresis. As a result, we suggest that the memory-conservation theory is a better place to start for the prediction of experimental measurables and the creation of device and circuit simulations. As this model is general and has been applied to other specific devices in~\cite{F1}, we would like to ask other experimentalists with memristors to also test this novel theory.

Unlike the phenomenological model, which is based on a phenomenological understanding of the memristor's operation, the memory-conservation theory is grounded in electrodynamics. As a result, the success of the memory-conservation theory shown here suggests that the model it is built on is more correct. This would require that we consider that Chua's constitutive relation between charge and magnetic flux refers to that associated with the oxygen vacancies. Thus, memristors will have to be considered as two level devices, whereby the vacancies' motion is experienced by the conducting electrons. (Note that the current state of the memory-conservation model only includes the electronic current as the circuit measurable $I$, we expect that inclusion of the vacancies as charge carriers will explain the non-zero current at zero voltage). 
 
If the memory-conservation model is considered verified, what does the existence of the memory conservation theory imply? Firstly, we must consider the concept of a magnetic field associated with ionic motion when talking about some electronic devices. Secondly, the theory helps to solve one of the problems in memristor physics, that of the non-existence of a relevant magnetic flux by pointing out which flux in the system is relevant. This allows us to tie Strukov's, our's and possibly many other workers' experimental memristors devices to Chua's elegant theory. Thirdly, the importance of tiny magnetic fields associated with ionic motion allows us to understand how memristor theory could be applied to neuronal operation. Although this has been suggested before~\cite{247,248}, we hope that the memory-conservation theory could be used to describe neural operation. Fourthly, there is the implications for resistance. The memory-conservation theory requires the, somewhat odd, novel understanding of resistance as 
being relative to the charge carrier experiencing it. This means that a material's resistance would be different dependent on the charge carrier passing through it (intuitively understood when we discuss inertia and ion mobilities in general). However, in electronics, as evidenced by the name of the field, we only care about the electron's experience of resistance. This work shows that, perhaps, we may need to collate tables of material data which record the resistance as felt by other charge carriers. It is our expectation that this would provide a new fabrication parameter to play around with and would aid the design of novel `electronic' materials.


\bibliography{UWELit}

\begin{thebibliography}{10}
\providecommand{\url}[1]{#1}
\csname url@samestyle\endcsname
\providecommand{\newblock}{\relax}
\providecommand{\bibinfo}[2]{#2}
\providecommand{\BIBentrySTDinterwordspacing}{\spaceskip=0pt\relax}
\providecommand{\BIBentryALTinterwordstretchfactor}{4}
\providecommand{\BIBentryALTinterwordspacing}{\spaceskip=\fontdimen2\font plus
\BIBentryALTinterwordstretchfactor\fontdimen3\font minus
  \fontdimen4\font\relax}
\providecommand{\BIBforeignlanguage}[2]{{%
\expandafter\ifx\csname l@#1\endcsname\relax
\typeout{** WARNING: IEEEtran.bst: No hyphenation pattern has been}%
\typeout{** loaded for the language `#1'. Using the pattern for}%
\typeout{** the default language instead.}%
\else
\language=\csname l@#1\endcsname
\fi
#2}}
\providecommand{\BIBdecl}{\relax}
\BIBdecl

\bibitem{15}
D.~B. Strukov, G.~S. Snider, D.~R. Stewart, and R.~S. Williams, ``The missing
  memristor found,'' \emph{Nature}, vol. 453, pp. 80--83, 2008.

\bibitem{224}
P.~Georgiou, S.~Yaliraki, E.~Drakakis, and M.~Barahona, ``Quantitative measure
  of hysteresis for memristors through explicit dynamics,'' \emph{Proceedings
  of the Royal Society A}, vol. 468, pp. 2210--2229, August 2012.

\bibitem{F1}
E.~Gale, ``The missing memristor magnetic flux found,'' \emph{Submitted,
  pre-print at http://arxiv.org/abs/1106.3170v2}, Submitted 2013.

\bibitem{14}
L.~O. Chua, ``Memristor - the missing circuit element,'' \emph{IEEE Trans.
  Circuit Theory}, vol.~18, pp. 507--519, 1971.

\bibitem{Chua1969}
------, \emph{Introduction to Nonlinear Network Theory}, 1st~ed.\hskip 1em plus
  0.5em minus 0.4em\relax McGraw-Hill; 1ST edition (1969), 1969.

\bibitem{279}
S.~P. Adhikari, H.~K. Maheshwar Pd.~Sah, and L.~O. Chua, ``Three fingerprints
  of memristor,'' \emph{IEEE Transactions on Circuits and Systems}, vol.~60,
  pp. 3008--3021, November 2013.

\bibitem{93}
Y.~V. Pershin and M.~D. Ventra, ``Memory effects in complex materials and
  nanoscale systems,'' \emph{Advances in Physics}, vol.~60, pp. 145--227, 2011.

\bibitem{276}
C.~L. S.~Grimes and O.~Martinsen, ``Memristive properties of human sweat
  ducts,'' \emph{World Congress on Medical Physics and Biomedial Engineering},
  vol. 25/7, pp. 696--698, 2009.

\bibitem{119}
L.~Chua, ``Resistance switching memories are memristors,'' \emph{Applied
  Physics A: Materials Science \& Processing}, pp. 765--782, 2011.

\bibitem{M1}
E.~Gale, D.~Pearson, S.~Kitson, A.~Adamatzky, and B.~de~Lacy~Costello, ``The
  effect of changing electrode metal on solution-processed flexible titanium
  dioxide memristors,'' \emph{Submitted}, 2013.

\bibitem{28}
N.~Gergel-Hackett, B.~Hamadani, B.~Dunlap, J.~Suehle, C.~Richer, C.~Hacker, and
  D.~Gundlach, ``A flexible solution-processed memrister,'' \emph{IEEE Electron
  Device Letters}, vol.~30, pp. 706--708, 2009.

\bibitem{71}
S.~H. Jo, T.~Chang, I.~Ebong, B.~B. Bhadviya, P.~Mazumder, and W.~Lu,
  ``Nanoscale memristor device as a synapse in neuromorphic systems,''
  \emph{Nanoletters}, vol.~10, pp. 1297--1301, 2010.

\bibitem{5}
V.~Erokhin and M.~Fontana, ``Electrochemically controlled polymeric device: a
  memristor (and more) found two years ago,'' \emph{arXiv:0807.0333v1
  [cond-mat.soft]}, 2008.

\bibitem{88}
A.~Smerieri, T.~Berzina, V.~Erokhin, and M.~Fontana, ``A functional polymeric
  material based on hybrid electrochemically controlled junctions,''
  \emph{Materials Science and Engineering C}, vol.~28, pp. 18--22, 2008.

\bibitem{94}
J.~Wu and R.~L. McCreery, ``Solid-state electrochemistry in molecule/tio2
  molecular heterojunctions as the basis of the tio2 ``memristor'',''
  \emph{Journal of the Electrochemical Society}, vol. 156, pp. P29--P37, 2009.

\bibitem{165}
D.-H. Kwon, K.~M. Kim, J.~H. Jang, J.~M. Jeon, M.~H. Lee, G.~H. Kim, X.-S. Li,
  G.-S. Park, B.~Lee, S.~Han, M.~Kim, and C.~S. Hwang, ``Atomic structure of
  conducting nanofilaments in tio2 resistive switching memory,'' \emph{Nature
  Nanotechnology}, vol.~5, pp. 148--153, 2010.

\bibitem{155}
R.~Wasser and M.~Aono, ``Nanoionics-based resistive switching memories,''
  \emph{Nature Materials}, vol.~6, pp. 833--840, 2007.

\bibitem{136}
K.~Szot, M.~Rogala, W.~Speier, Z.~Klusek, A.~Besmehn, and R.~Waser, ``{TiO$_2$}
  - a prototypical memristive material,'' \emph{Nanotechnology}, vol.~22, pp.
  254\,001--1 -- 254\,001--21, June 2011.

\bibitem{154}
A.~B. S.E.~Savel'ev, A.S.~Alexandrov and R.~S. Williams, ``Molecular dynamics
  simulations of oxide memory resistors (memristors),'' \emph{Nanotechnology},
  vol.~22, p. 254011, 2011.

\bibitem{168}
D.~Ielmini, F.~Nardi, and C.~Cagli, ``Universal reset characteristics of
  unipolar and bipolar metal-oxide rram,'' \emph{IEEE Transactions on Electron
  Devices}, vol.~58, pp. 3246--3253, 2011.

\bibitem{RevMemReRAM}
E.~Gale, ``Memristors and reram: Materials, mechanisms and models (a review),''
  \emph{Submitted}.

\bibitem{289}
S.~Williams, Private Communication.

\bibitem{63}
D.~Biolek, Z.~Biolek, and V.~Biolkova, ``Spice modeling of memristive,
  memcapacitative and meminductive systems,'' \emph{European Conference on
  Circuit Theory Design}, pp. 249--252, 2009.

\bibitem{65}
M.~Stork, J.~Hrusak, and D.~Mayer, ``Memristor based feedback systems,'' in
  \emph{Applied Electronics, 2009. AE 2009}, 2009, pp. 237--240.

\bibitem{66}
Y.~Z.~X. Zhang and J.~Yu, ``Approximated spice model for memristor,''
  \emph{2009 INTERNATIONAL CONFERENCE ON COMMUNICATIONS, CIRCUITS AND SYSTEMS
  PROCEEDINGS, VOLUMES I \& II}, pp. 928--931, 2009.

\bibitem{73}
K.~Witrisal, ``Memristor-based stored-reference receiver - the uwb solution,''
  \emph{Electronics Letters}, vol.~45, 2009.

\bibitem{78}
K.~Eshraghian, K.-R. Cho, O.~Kavehei, S.-K. Kang, D.~Abbott, and S.-M.~S. Kang,
  ``Memristor mos content addressable memory (mcam): Hybrid architecture for
  future high performance search engines,'' \emph{Very Large Scale Integration
  (VLSI) Systems, IEEE Transactions on}, vol.~19, no.~8, pp. 1407--1417, 2011.

\bibitem{83}
X.~Zhang, Z.~Huang, and J.~Yu, ``Memristor model for spice,'' \emph{IEICE
  Trans. Electron.}, vol. E93-C, pp. 355--360, 2010.

\bibitem{236}
C.~T. Themistoklis~Prodromakis and L.~Chua, ``Two centuries of memristors,''
  \emph{Nature Materials}, vol.~11, pp. 478--481, 2012.

\bibitem{280}
J.-M. Ginoux and B.~Rossetto, \emph{The Singing Arc: The Oldest Memristor?},
  A.~Adamatzky and G.~Chen, Eds.\hskip 1em plus 0.5em minus 0.4em\relax World
  Scientific, 2012.

\bibitem{29}
B.~Widrow, ``An adaptive 'adaline' neuron using chemical 'memistors',''
  \emph{Technical Report}, 1960.

\bibitem{M0}
E.~Gale, R.~Mayne, A.~Adamatzky, and B.~de~Lacy~Costello, ``Drop-coated
  titanium dioxide memristors,'' \emph{Materials Chemistry and Physics}, vol.
  143, pp. 524--529, January 2014.

\bibitem{WasersBook}
R.~Waser, Ed., \emph{Nanoelectronics and Information Technology}, 3rd~ed.\hskip
  1em plus 0.5em minus 0.4em\relax Wiley-VCH, 2002.

\bibitem{254}
E.~M. Gale, B.~{de Lacy Costello}, and A.~Adamatzky, ``Filamentary extension of
  the {Mem-Con} theory of memristance and its application to titanium dioxide
  {Sol-Gel} memristors,'' in \emph{2012 IEEE International Conference on
  Electronics Design, Systems and Applications (ICEDSA 2012)}, Kuala Lumpur,
  Malaysia, Nov. 2012.

\bibitem{SpcJ}
E.~Gale, B.~de~Lacy~Costello, and A.~Adamatzky, ``Observation, characterization
  and modeling of memristor current spikes,'' \emph{Appl. Math. Inf. Sci.},
  vol.~7, pp. 1395--1403, 4,July 2013.

\bibitem{232}
B.~Bo-Cheng, X.~J. Ping, Z.~Guo-Hua, M.~Zheng-Hua, and Z.~Ling, ``Chaotic
  memristive circuit: equivalent circuit realization and dynamical analysis,''
  \emph{Chinese Physics B}, vol.~20, p. 120502 (7pp), 2011.

\bibitem{247}
L.~Chua, V.~Sbitnev, and H.~Kim, ``Hodgkin-huxley axon is made of memristors,''
  \emph{International Journal of Bifurcation and Chaos}, vol.~22, p. 1230011
  (48pp), 2012.

\bibitem{248}
------, ``Neurons are poised near the edge of chaos,'' \emph{International
  Journal of Bifurcation and Chaos}, vol.~11, p. 1250098 (49pp), 2012.

\end{thebibliography}

\end{document}